\documentclass[11pt]{article}

\usepackage[usenames,dvipsnames]{pstricks}
\usepackage{epsfig}
\usepackage{pst-grad} 
\usepackage[space]{grffile} 
\usepackage{etoolbox} 
\makeatletter 
\patchcmd\Gread@eps{\@inputcheck#1 }{\@inputcheck"#1"\relax}{}{}
\makeatother

\usepackage{graphicx}
\usepackage{amsmath}
\usepackage{psfrag}
\usepackage{color,soul}
\usepackage{natbib}
\usepackage{siunitx}
\usepackage{float}
\usepackage{overpic}
\usepackage{tikz}
\usepackage{multirow}
\usepackage{bm}
\usepackage{textgreek}
\usepackage{url}
\usepackage{amssymb}
\usepackage{wasysym}

\usepackage{xfrac}
\usepackage{setspace}
\usepackage{gensymb}
\usepackage{mathtools}

\usepackage{arydshln}

\usetikzlibrary{arrows}

\usepackage[noblocks]{authblk}
\usepackage[margin=1in]{geometry}

\usepackage[title]{appendix}

  \DeclareTextFontCommand\textsfi{\usefont{OT1}{cmss}{m}{sl}}
  \DeclareMathAlphabet\mathsfi            {OT1}{cmss}{m}{sl}
  \DeclareTextFontCommand\textsfb{\usefont{OT1}{cmss}{bx}{n}}
  \DeclareMathAlphabet\mathsfb            {OT1}{cmss}{bx}{n}
  \DeclareTextFontCommand\textsfbi{\usefont{OT1}{cmss}{m}{sl}}
  \DeclareMathAlphabet\mathsfbi            {OT1}{cmss}{m}{sl}
  \IfFileExists{t1phv.fd}
  {\DeclareTextFontCommand\textsfbi{\usefont{T1}{phv}{b}{it}}
  \DeclareMathAlphabet\mathsfbi            {T1}{phv}{b}{it}
  }{}
  \IfFileExists{ot1phv.fd}
  {\DeclareTextFontCommand\textsfbi{\usefont{OT1}{phv}{b}{it}}
  \DeclareMathAlphabet\mathsfbi            {OT1}{phv}{b}{it}
  }{}

\newcommand\affiliation[1]{\gdef\@affiliation{\let\aff\aff@inst#1}}
\gdef\@affiliation{}

\def\email#1{Email address for correspondence: #1}
\def\aff#1{\ignorespaces\textsuperscript{#1}}
\def\corresp#1{\unskip\thanks{#1}}

\numberwithin{equation}{section}

\renewenvironment{abstract}
{\begin{quote}
\noindent \rule{\linewidth}{.5pt}\par{\bfseries \abstractname.}}
{\medskip\noindent \rule{\linewidth}{.5pt}
\end{quote}
}

\newcommand{\Eqn}[1]{Equation~(\ref{Eq:#1})}

\newcommand{\partderiv}[2]{\frac{\partial #1}{\partial #2}}

\newcommand{\tc}[1]{\mathsfbi{#1}}	

\DeclareMathOperator{\ImV}{Im}

\DeclareSymbolFont{matha}{OML}{txmi}{m}{it}
\DeclareMathSymbol{\varv}{\mathord}{matha}{118}

\newcommand{\cmmnt}[1]{\ignorespaces}

\definecolor{blue}{rgb}{0,0,1}
\definecolor{blue2}{rgb}{0,0.502,1}
\definecolor{red}{rgb}{0.8,0,0}
\definecolor{red2}{rgb}{1,0,0}
\definecolor{green}{rgb}{0,1,0}
\definecolor{green2}{rgb}{0.2353,0.6275,0.2353}
\definecolor{magenta}{rgb}{1,0.2,1}
\definecolor{orange}{rgb}{1,0.5,0}
\definecolor{orange2}{rgb}{0.9686,0.6353,0.3020}
\definecolor{black}{rgb}{0,0,0}
\definecolor{gray}{rgb}{0.85,0.85,0.85}
\definecolor{gray2}{rgb}{0.4,0.4,0.4}
\definecolor{magenta}{rgb}{1,0,1}
\definecolor{yellow}{rgb}{0.749,0.749,0}
\definecolor{teal}{rgb}{0,0.749,0.749}

\usepackage{hyperref}

\definecolor{darkblue}{rgb}{0,0,0.80}

\hypersetup{
    colorlinks=true,
    linkcolor={darkblue},
    citecolor={darkblue},
    filecolor=black,
    urlcolor={darkblue},
    plainpages=false,
}

\title{\bf Efficient global resolvent analysis via the one-way Navier-Stokes equations.  Part 1.  Forced response}

\author[1]{\bf Aaron Towne\corresp{\email{towne@umich.edu}}}
\author[2]{\bf Georgios Rigas}
\author[3]{\bf Ethan Pickering}
\author[3]{\bf Tim Colonius}

\affil[1]{\normalsize Department of Mechanical Engineering, University of Michigan, Ann Arbor, MI, USA  }
\affil[2]{\normalsize Imperial College London, South Kensington, London, United Kingdom}
\affil[3]{\normalsize California Institute of Technology, Pasadena, CA, USA \vspace{-1cm}}

\date{}

\begin{document}
\maketitle

\begin{abstract}
Resolvent analysis is a powerful tool for modeling and analyzing turbulent flows and in particular provides an approximation of coherent flow structures.  Despite recent algorithmic advances, computing resolvent modes for flows with more than one inhomogeneous spatial coordinate remains computationally expensive.  In this two-part paper, we show how efficient and accurate approximations of resolvent modes can be obtained using a well-posed spatial marching method for flows that contain a slowly varying direction.  In this first part of the paper, we derive a well-posed and convergent one-way equation describing the downstream-traveling waves supported by the linearized Navier-Stokes equations.  Integrating these equations, which requires significantly less CPU and memory resources than a direct solution of the linearized Navier-Stokes equations, approximates the action of the resolvent operator on a forcing vector.  This capability is leveraged in part 2 of the paper to compute approximate resolvent modes.  The method is validated and demonstrated using the examples of a simple acoustics problem and a supersonic turbulent jet. \\
\end{abstract}



\section{Introduction}
\label{Sec:Intro}

Resolvent analysis (also called input/output analysis or frequency response analysis) is a powerful and popular tool for studying linear energy amplification mechanism within the Navier-Stokes equations. The resolvent operator is derived from the linearized Navier-Stokes (LNS) equations and constitutes a transfer function between inputs and outputs of interest. It has been used to study the linear response of flows to external excitation \citep{Trefethen:1993, Farrell:2001, Jovanovic:2005, Sipp:2010} and to forcing from the nonlinear terms in the Navier–Stokes equations \citep{McKeon:2010}. In the latter context, the method can be derived by reorganizing the Navier–Stokes equations into terms that are linear and nonlinear with respect to perturbations to the turbulent mean flow.  The singular value decomposition (SVD) of the resolvent operator associated with the linearized Navier-Stokes equations then identifies modes that are optimal in terms of their linear gain between the nonlinear terms and the perturbations to the mean.  These optimal modes have been shown to provide a useful model of coherent structures within turbulent flows, and in particular provide an approximation of the space-time coherent structures educed from data using spectral proper orthogonal decomposition \citep{Towne2018spectral}.

The computational resources required to compute resolvent modes depends critically on the number of spatial dimensions that must be numerically discretized.  The linearized Navier-Stokes equations nominally contain three spatial dimensions, but the equations can be simplified by expanding the flow variables into Fourier modes in homogenous dimensions, i.e., directions in which the base flow (about which the equations are linearized) does not vary.  This drastically reduces the size of the discretized operators that must be manipulated, decreasing the computational cost of the model.  Methods in which all inhomogeneous dimensions are discretized are often called global methods \citep{Theofilis2011global}.  

Computing global resolvent modes for flows with more than one inhomogeneous direction remains a computationally intensive task.  Whereas flows with one inhomogeneous direction can now be tackled on a lap top computer, high-memory workstations and large-scale clusters must be employed for flows with two and three inhomogeneous directions, respectively.  These requirements limit the utility of global resolvent analysis for many flows of interest and mitigate some of their advantages compared to fully nonlinear numerical simulations.

Recent efforts to reduce the cost of resolvent analysis have focused on reducing the cost of computing the SVD of the resolvent operator.  For example, \cite{Moarref:2014} used a randomized singular value decomposition algorithm (RSVD) to compute resolvent modes for a turbulent channel flow (which has only one inhomogeneous direction) and reported that this reduced the cost of the calculations by a factor of two.  \cite{Ribeiro2020randomized} offered several improvements for the application of RSVD to resolvent analysis and achieved an order-of-magnitude speedup compared to standard SVD algorithms for a separated flow around an airfoil (which has two inhomogesous directions).  Given the reduced cost of computing the SVD, the majority of the remaining cost in their algorithm is associated with computing the action of the resolvent operator on a vector.  This is true also for time-stepping methods \citep{Monokrousos2010global, Martini2021efficient, Farghadan2021randomized}, where the dominant cost is integrating direct and adjoint linear equation in the time domain used to apply the resolvent operator and its adjoint.

In this two-part paper, we develop a method that significantly reduces the cost of computing resolvent modes for flows that include a slowly varying spatial direction, i.e., a direction in which the mean flow is inhomogeneous but changes gradually.  This common class of flows includes free-shear flows like mixing layers and jets as well as wall-bounded flows with spatially developing boundary layers on gradually changing objects like a flat plate, cone, airfoil, etc.

In part 1 of the paper, we show how the action of the resolvent operator on a forcing vector can be efficiently and accurately approximated for slowly varying flows using a well-posed spatial marching technique. In  Part 2 \citep{Rigas2020fast},  we show how this capability can be used to compute the singular modes of the resolvent operator using iterative downstream and upstream marching.

Spatial marching methods are commonly applied to slowly varying flows in order to compute approximate eigenmodes or the downstream response to an initial disturbance introduced at some upstream location.  The classical tool for these purposes is the parabolized stability equations (PSE), which constitute an ad hoc generalization of classical parallel flow stability theory \citep{Bertolotti:1992, Herbert:1997}.  The basic idea of PSE is to separate the flow variables at each frequency into a slowly varying shape function and a rapidly varying wave-like component.  Inserting this ansatz into the linearized Navier-Stokes equations leads to a modified set of equations for the shape function, which can be rapidly solved via spatial integration in the slowly varying direction, leading to an approximation of the downstream response to an initial disturbance.  The initial disturbance is usually chosen to be a locally parallel eigenmode, in which case the PSE solution is interpreted as a weakly nonparallel eigenmode of the flow.  

Several authors have recently observed that for flows dominated by a single convective instability, the PSE solution appears to provide a reasonable approximation of the leading resolvent output mode, i.e., the left singular vector of the resolvent operator with the largest singular value \citep{Jeun:2016, Beneddine:2016}.  However, there are several issues that limit the utility of PSE for approximating resolvent modes.  First, PSE does not provide any information about the input resolvent modes or gains, both of which are critical for analysis and modeling.  Second, by construction, PSE is incapable of computing sub-optimal modes, i.e., it can only approximate one mode per frequency.  Third, PSE can accurately capture the influence of only a single instability mechanism.  This limitation stems from the fact that, despite their name, the parabolized stability equations are, in fact, elliptic in the slowly varying direction due to the boundary value nature of the linearized Navier-Stokes equations \citep{Li:1997}.  As a result, the PSE spatial integration is mathematically ill-posed, and regularization methods are required to stabilize the spatial march.  Recently, \cite{Towne2019critical} showed that these regularization methods contaminate the PSE solution, except in cases where the flow is dominated by a single instability mode at each frequency.  As a result, PSE is not an appropriate tool for flows in which multiple modes are of interest, including multiple instability mechanisms, transient growth, or acoustics.

\begin{figure}
\centering
 \includegraphics[width=0.9\linewidth,trim={0cm 0cm 0cm 0},clip]{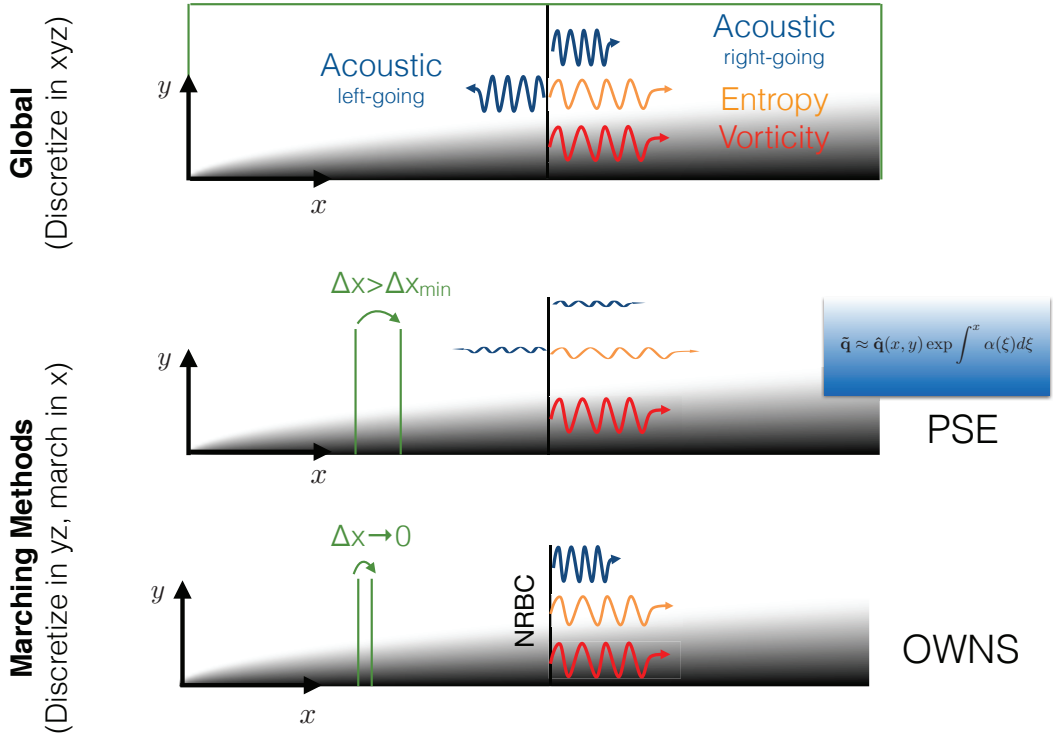}
 \caption{Global versus spatial marching problems.}
 \label{fig:OWNScartoon}
\end{figure}

\cite{Towne2015oneway} introduced an alternative method for obtaining fast, approximate linear solutions for slowly varying flows that overcomes these issues by constructing well-posed spatial evolution equations that do not require detrimental PSE-like regularization.  Using ideas originally developed for constructing high-order non-reflecting boundary conditions \citep{Hagstrom:2004}, the flow variables are decomposed into upstream and downstream propagating waves in the slowly varying direction.  An approximate evolution equation is derived for the downstream waves, which can be solved via a well-posed spatial march.  The method, which has come to be known as the one-way Navier-Stokes equations (OWNS), has been applied to both free-shear flows such as mixing layers \citep{Towne:2013} and jets \citep{Towne:2014, Rigas:2017b} as well as wall-bounded flows \citep{Rigas:2017a, Kamal2020application} and is typically more than an order-of-magnitude faster than global methods.  A schematic comparison of global methods, PSE, and OWNS is provided in figure~\ref{fig:OWNScartoon}.  Unfortunately, the original formulation of OWNS developed by \cite{Towne2015oneway} cannot accommodate a forcing term on the linearized flow equations.  Such a forcing term is fundamental to resolvent analysis, making this formulation unsuitable for computing resolvent modes.

To enable the efficient computation of resolvent modes using spatial marching, we introduce a new variant of OWNS that naturally accommodates a forcing terms.  The method is formulated in terms of a projection operator that splits the flow variables into upstream and downstream traveling components and which can be applied to the linearized Navier-Stokes equations to obtain evolution equations for each set of waves.  Importantly, the projection operator can also be used to split arbitrary forcing terms into parts which influence downstream and upstream traveling waves, enabling its use for approximating resolvent modes.  

To distinguish between the two variants of OWNS, we will refer to the original formulation as OWNS-O and the new formulation as OWNS-P, in recognition of their connections with outflow boundary conditions and a projection operator, respectively. 
 
The remainder of this paper is organized as follows.  The OWNS-P method is formulated and mathematically analyzed in \S~\ref{Sec:Formulation}.  It is then demonstrated in \S~\ref{Sec:Examples} using two example problems – a simple acoustic wave propagation problem and a turbulent jet.  Finally, the paper is summarized in \S~\ref{Sec:Conclusions}.


\section{One-way Navier-Stokes equations -- a projection approach}
\label{Sec:Formulation}

\subsection{Problem setup}

We begin with the compressible Navier-Stokes equations in an arbitrary orthogonal coordinate systems, written abstractly as 
\begin{equation}
\label{Eq:NavierStokes}
\frac{\partial q}{\partial t} = \mathcal{N}(q).
\end{equation}
Equation~(\ref{Eq:NavierStokes}) contains mass, momentum, and energy equations, and the state vector $q(x,y,z,t)$ contains an appropriate set of variables, e.g., velocity components and two thermodynamic variables such as density and pressure.  

Applying the Reynolds decomposition
\begin{equation}
\label{Eq:ReynoldsDecomp}
q(x,y,z,t) = \bar{q}(x,y,z) + q^{\prime}(x,y,z,t)
\end{equation}
and moving terms that are linear and nonlinear in the fluctuation $q^{\prime}$ to the left- and right-hand-sides of the equations, respectively, leads to an equation of the form
\begin{equation}
\label{Eq:LinearPDE_L}
\partderiv{q^{\prime}}{t} + \mathcal{L} q^{\prime} = f.
\end{equation}
The left-hand-side of (\ref{Eq:LinearPDE_L}) is the linearized Navier-Stokes equations, while the right-hand-side vector $f(x,y,z,t)$ contains all remaining nonlinear terms, which within the context of resolvent analysis are interpreted as an external forcing on the linearized equations.

The relationship between the nonlinear term and the linear fluctuation can be expressed in the frequency domain as
\begin{equation}
\label{Eq:q_eq_Rf}    
\hat{q} = \mathcal{R}_{G} \hat{f},    
\end{equation}
where $\hat{q}$ and $\hat{f}$ are the Fourier transforms of $q^{\prime}$ and $f$, respectively, and
\begin{equation}
\label{Eq:ResolventOp}
\mathcal{R}_{G} = \left( i \omega \mathcal{I} + \mathcal{L} \right)^{-1}
\end{equation}
is the global resolvent operator.  Resolvent modes are given by the SVD of the resolvent operator
\begin{equation}
\label{Eq:ResolventSVD}
\mathcal{R}_{G} = U \Sigma V^{*}.
\end{equation}
The singular values, which appear within the diagonal positive-semi-definite matrix $\Sigma$, give the square root of the optimal gains between the input and output modes defined by the right and left singular vectors contained in the columns of the orthonormal matrices $V$ and $U$, respectively.

In practice, computing resolvent modes requires numerical discretization of~(\ref{Eq:LinearPDE_L}) in all inhomogeneous directions.  While the explicit construction of $\mathcal{R}_{G}$ can usually be avoided \citep{Jeun:2016, Schmidt2018spectral, Ribeiro2020randomized}, it is necessary to compute the action of the resolvent operator on an arbitrary forcing vector, i.e., to evaluate the discretized form of~(\ref{Eq:q_eq_Rf}).  In what follows, we will show how the slow variation of the mean flow present in many flows can be leveraged to efficiently and accurately approximate the action of the resolvent operator on a forcing vector using a spatial marching method.  Then, in part 2 of the paper \citep{Rigas2020fast}, we will show how this capability can be used to computer optimal forces and responses, i.e., the singular modes of the approximate resolvent operator.

\subsection{Identifying downstream- and upstream-traveling waves}
\label{Sec:RightVsLeft}


The first critical step in developing a well-posed one-way equation is to identify parts of the solution that transfer energy in the positive and negative streamwise directions, which we call downstream-traveling and upstream-traveling, respectively.  To this end, we rewrite~(\ref{Eq:LinearPDE_L}) as
\begin{equation}
\label{Eq:LinearPDE_f}
\partderiv{q^{\prime}}{t} + \mathcal{A} \partderiv{q^{\prime}}{x} + \mathcal{B}  q^{\prime} + \mathcal{C} \partderiv{q^{\prime}}{x} + \mathcal{D} \frac{\partial^2 q^{\prime}}{\partial x^2} = f.
\end{equation}
Here, $x \in \mathbb{R}$ is the slowly varying direction along which we will apply spatial marching, and $y$ and $z$ are additional, transverse spatial dimensions.  In~(\ref{Eq:LinearPDE_f}), we have isolated $x$-derivative terms arising from the convective terms in the Navier-Stokes equations ($\mathcal{A}\frac{\partial}{\partial x}$) and from streamwise viscous terms ($\mathcal{C}\partderiv{}{x}$, $\mathcal{D} \frac{\partial^2 }{\partial x^2}$); the linear operator $\mathcal{B}$ contains all other terms in the linearized Navier-Stokes equations.  

To obtain a one-way equation, we wish to work in terms of a system including only first $x$-derivatives \citep{Towne2015oneway}.  This can be accomplished in one of several ways.  The viscous terms can be parabolized by redefining the state vector to include both $q^{\prime}$ and $\partderiv{q^{\prime}}{x}$ and writing~(\ref{Eq:LinearPDE_f}) as an expanded system (of twice the orignal size) using this new state variable \citep{Towne:2016, Harris2020well}, analogous to the standard approach for solving quadratic eigenvalue problems \citep{Tisseur2001quadratic}.  Alternatively, the streamwise viscous terms can be moved to the right-hand-side of~(\ref{Eq:LinearPDE_f}) and treated as a forcing term, which is later evaluated using the solution at the previous step in the spatial march \citep{Kamal2020application}.  Finally, following the standard boundary-layer approximation, the streamwise viscous terms can simply be neglected.  We have found this simplification to be sufficient for all flows, including both free-shear and wall-bounded flows, to which OWNS has been applied to date.  Accordingly, we neglect $\mathcal{C}\partderiv{}{x}$ and $\mathcal{D} \frac{\partial^2 }{\partial x^2}$ in what follows.

Next, we discretize (\ref{Eq:LinearPDE_f}) in the transverse directions using a collocation method (such as finite differences) with  $N_{c}$ collocation points.  The semi-discrete approximation of (\ref{Eq:LinearPDE_f}) can then be written
\begin{equation}
\label{Eq:SemiDiscPDE}
\partderiv{\boldsymbol{q}^{\prime}}{t} + \tc{A}(x) \partderiv{\boldsymbol{q}^{\prime}}{x} + \tc{B} (x)\boldsymbol{q}^{\prime} = \boldsymbol{f}(x,t),
\end{equation}
where $\boldsymbol{q}^{\prime}(x,t), \boldsymbol{f}(x,t) \in \mathbb{C}^{N}$ and $\tc{A}, \tc{B} \in \mathbb{C}^{N \times N}$ are semi-discrete analogues of $q$, $f$, $\mathcal{A}$ and $\mathcal{B}$, respectively.  The entries of $\tc{A}$ consist of the values of $\mathcal{A}$ at the collocation points, while the matrix $\tc{B}$ contains discrete approximations of transverse derivatives contained within $\mathcal{B}$ as well as modifications required to enforce the desired transverse boundary conditions.   The total size of the semi-discrete system is $N = N_{q}N_{c}$, where $N_{q}$ is the number of state variables in the Navier-Stokes equations (e.g., $N_{q} = 5$ for three dimensional problems).



\Eqn{SemiDiscPDE} is a one-dimensional strongly 
hyperbolic system since $\tc{A}$ is diagonalizable and has real eigenvalues.  That is, there exists a transformation $\tc{T}(x)$ such that  
\begin{equation}
\label{Eq:charTrans}
\tc{T}\tc{A}\tc{T}^{-1} = \tilde{\tc{A}} = \left[  \begin{array}{ccc} \tilde{\tc{A}}_{++} & \tc{0} & \tc{0} \\ \tc{0} & \tilde{\tc{A}}_{--} & \tc{0} \\ \tc{0} & \tc{0} & \tilde{\tc{A}}_{00} \end{array} \right],
\end{equation}
where $\tilde{\tc{A}}$ is a diagonal matrix and the diagonal entries of the sub-matrices $\tilde{\tc{A}}_{++} \in \mathbb{R}^{N_{+} \times N_{+}} > 0$,  $\tilde{\tc{A}}_{--} \in \mathbb{R}^{N_{-} \times N_{-}} < 0$, and $\tilde{\tc{A}}_{00} \in \mathbb{R}^{N_{0} \times N_{0}} = 0$ contain the positive, negative, and zero eigenvalues of $\tc{A}$, respectively.  Here, $N_{+}$, $N_{-}$, and $N_{0}$ denote the number of positive, negative, and zero eigenvalues, respectively, and $N = N_{+} + N_{-} + N_{0}$.  The transformation $\tc{T}$ is known analytically since it is the discretization of the matrix that diagonalizes $\mathcal{A}$.


To derive a one-way equation, it is convenient to work in terms of the characteristic variables of (\ref{Eq:SemiDiscPDE}), which are defined in terms of the transformation $\tc{T}(x)$,
\begin{equation}
\label{Eq:CharVars}
\boldsymbol{\phi}(x,t) = \tc{T}(x)\boldsymbol{q}^{\prime}(x,t).
\end{equation}
These characteristic variables can be split into three components associated with the positive, negative, a zero blocks of $\tilde{\tc{A}}$, 
\begin{equation}
\label{eq:charVars_split_pm}
\boldsymbol{\phi} = \left\{\begin{array}{c} \boldsymbol{\phi}_{+} \\ \boldsymbol{\phi}_{-} \\ \boldsymbol{\phi}_{0}  \end{array}\right\}
\end{equation}
with $\boldsymbol{\phi}_{+} \in \mathbb{R}^{N_{+}}$,  $\boldsymbol{\phi}_{-} \in \mathbb{R}^{N_{-}}$, and $\boldsymbol{\phi}_{0} \in \mathbb{R}^{N_{0}}$.  For later use, we also define
\begin{equation}
\label{Eq:charVars_split_pm}
\boldsymbol{\phi}_{\pm} = \left\{\begin{array}{c} \boldsymbol{\phi}_{+} \\ \boldsymbol{\phi}_{-}  \end{array}\right\},
\end{equation}
which contains only the characteristic variables associated with the nonzero block of $\tilde{\tc{A}}$,
\begin{equation}
\label{Eq:A_tilde_pm}
\tilde{\tc{A}}_{\pm\pm} = \left[  \begin{array}{cc} \tilde{\tc{A}}_{++} & \tc{0} \\ \tc{0} & \tilde{\tc{A}}_{--} \end{array} \right].
\end{equation}
To be clear, the $\pm$ subscript here and throughout the paper does not indicate that we are choosing either the plus or minus characteristic, as in the typical usage of that symbol, but rather both component, as exemplified in~(\ref{Eq:charVars_split_pm}) and~(\ref{Eq:A_tilde_pm}).

In terms of the characteristic variables, (\ref{Eq:SemiDiscPDE}) becomes
\begin{equation}
\label{Eq:SemiDiscChar}
\partderiv{\boldsymbol{\phi}}{t} + \tilde{\tc{A}}\left(x\right)\partderiv{\boldsymbol{\phi}}{x} + \tilde{\tc{B}}\left(x\right)\boldsymbol{\phi} = \boldsymbol{f}_{\phi}(x,t),
\end{equation}
where $\tilde{\tc{B}} = \tc{T}\tc{B}\tc{T}^{-1} + \tilde{\tc{A}}\tc{T}\frac{d\tc{T}^{-1}}{dx}$ and $\boldsymbol{f}_{\phi} = \tc{T} \boldsymbol{f}$.

We ultimately wish to obtain the response to a forcing in the frequency domain.  However, we proceed by applying a Laplace transform in time, rather than a Fourier transform, to (\ref{Eq:SemiDiscChar}), giving
\begin{equation}
\label{Eq:SemiDiscFT}
s \, \boldsymbol{\hat{\phi}} + \tilde{\tc{A}}\left(x\right) \frac{d \boldsymbol{\hat{\phi}}}{dx} + \tilde{\tc{B}} \left(x\right)\boldsymbol{\hat{\phi}} = \boldsymbol{\hat{f}}_{\phi}(x,s),
\end{equation}
where $\boldsymbol{\hat{\phi}}(x,s)$ is the Laplace transform of $\boldsymbol{\phi}(x,t)$ and $s = \eta - i \omega$ ($\eta, \omega \in \mathbb{R}$) is the Laplace dual of $t$.  We will ultimately take $\eta = 0$ and set $\omega$ to a particular value to obtain the response to a forcing at that frequency, but keeping the possibility of non-zero $\eta$ will help us distinguish between upstream- and downstream-travelling solutions of (\ref{Eq:SemiDiscChar}).

Up to this point, we have made no approximation (aside from potentially discarding a small subset of viscous terms), and the action of the resolvent operator on a forcing vector could be computed by discretizing~(\ref{Eq:SemiDiscFT}) in $x$, applying boundary conditions at the beginning and end of the $x$ domain, solving for $\boldsymbol{\hat{\phi}}$, and inverting the transformation~(\ref{Eq:CharVars}) to obtain $\boldsymbol{\hat{q}}$.  However, this involves solving a large system of equations, as discussed in \S~\ref{Sec:Formulation}, which constitutes a large fraction of the cost of resolvent analysis.  Instead, we will obtain an approximate solution of~(\ref{Eq:SemiDiscFT}) via spatial integration.  Directly integrating~(\ref{Eq:SemiDiscFT}) is ill-posed \citep{Li:1997, Towne2015oneway, Towne2019critical}, leading to exponential divergence of the solution.  To circumvent this, we will derive a well-posed one-way equation that can be stably integrated.

To proceed, we isolate the $x$-derivatives within (\ref{Eq:SemiDiscFT}), giving
\begin{subequations}
\label{Eq:DAE}
\begin{align}
\tilde{\tc{A}}_{\pm\pm} \frac{d \boldsymbol{\hat{\phi}}_{\pm}}{dx} &= \tc{L}_{\pm\pm} \boldsymbol{\hat{\phi}}_{\pm} + \tc{L}_{\pm0} \boldsymbol{\hat{\phi}}_{0} + \boldsymbol{\hat{f}}_{\phi,\pm}, \label{Eq:DAE_a} \\
\boldsymbol{0} &= \tc{L}_{0\pm} \boldsymbol{\hat{\phi}}_{\pm} + \tc{L}_{00} \boldsymbol{\hat{\phi}}_{0} + \boldsymbol{\hat{f}}_{\phi,0}  \label{Eq:DAE_b}
\end{align}
\end{subequations}
with 
\begin{equation}
\label{Eq:defL}
\tc{L}(x,s) = \left[  \begin{array}{cc} {\tc{L}}_{\pm\pm} & {\tc{L}}_{\pm0}  \\ {\tc{L}}_{0\pm} & {\tc{L}}_{00} \end{array} \right] = - \left(s\tc{I} + {\tc{B}}\left(x\right) \right).
\end{equation}
Here and throughout the paper, the subscripts of a matrix indicate its size, e.g., ${\tc{L}}_{0\pm} \in \mathbb{C}^{N_0 \times (N_{+}+N_{-})}$.

Equation~(\ref{Eq:DAE}) is a differential-algebraic equation (DAE) due to the zero left-hand-side of~(\ref{Eq:DAE_b}), which occurs because of the zero eigenvalues of $\tc{A}$ contained in the zero matrix $\tilde{\tc{A}}_{00}$.  These zero eigenvalues correspond to points in the base flow where the streamwise velocity is zero or exactly sonic \citep{Towne2015oneway}.  The algebraic conditoins in~(\ref{Eq:DAE_b}) function as a constraint on the allowable form of $\boldsymbol{\hat{\phi}}$.  Assuming that ${\tc{L}}_{00}$ is invertible, then~(\ref{Eq:DAE}) is an DAE of index 1 and the zero characteristic variable $\boldsymbol{\hat{\phi}}_{0}$ is slaved to positive and negative characteristics variables as
\begin{equation}
\label{Eq:phi0_slaved}
\boldsymbol{\hat{\phi}}_{0} = - \tc{L}_{00}^{-1} \left(  \tc{L}_{0\pm} \boldsymbol{\hat{\phi}}_{\pm} + \boldsymbol{\hat{f}}_{\phi,0} \right).
\end{equation}

To obtain a one-way equation, it is convenient to reduce~(\ref{Eq:DAE}) to an ordinary differential equation (ODE) for $\boldsymbol{\hat{\phi}}_{\pm}$, which is accomplished by using~(\ref{Eq:phi0_slaved}) to eliminate $\boldsymbol{\hat{\phi}}_{0}$ from~(\ref{Eq:DAE_a}), leading to 
\begin{equation}
\label{Eq:EllipticBVP}
\frac{d \boldsymbol{\hat{\phi}}_{\pm}}{dx} = \tc{M}(x, s) \boldsymbol{\hat{\phi}}_{\pm} + \boldsymbol{\hat{h}}(x,s)
\end{equation}
with 
\begin{equation}
\label{Eq:defM}
\tc{M} = \tc{A}_{\pm\pm}^{-1} \left( \tc{L}_{\pm\pm} - \tc{L}_{\pm0} \tc{L}_{00}^{-1} \tc{L}_{0\pm} \right)
\end{equation}
and
\begin{equation}
\label{Eq:def_h}
\boldsymbol{\hat{h}} = \tilde{\tc{A}}_{\pm\pm}^{-1} \left( \boldsymbol{\hat{f}}_{\phi,\pm} - \tc{L}_{\pm0} \tc{L}_{00}^{-1} \boldsymbol{\hat{f}}_{\phi,0} \right).
\end{equation}
While these expressions formally contain $\tc{L}_{00}^{-1}$, this inverse, and its potential detrimental effect on the sparsity of $\tc{M}$, is in practice avoided by reversing the contraction of the system (from~(\ref{Eq:DAE}) to~(\ref{Eq:EllipticBVP})) in the final set of one-way equations (see \S~\ref{sec:implemenation} and Appendix~\ref{App:OWNS_mats}).

If $N_0 = 0$, which is the case, e.g., in subsoninc free-shear flows, then all matrices containing a zero index vanish, $\boldsymbol{\hat{\phi}} = \boldsymbol{\hat{\phi}}_{\pm}$, and (\ref{Eq:EllipticBVP}) and~(\ref{Eq:def_h}) reduce to the simpler forms
\begin{equation}
\label{Eq:defM_simple}
\tc{M} = -\tilde{\tc{A}}^{-1}\left(s\tc{I} + \tilde{\tc{B}} \right).
\end{equation}
and 
\begin{equation}
\label{Eq:def_h_simple}
\boldsymbol{\hat{h}} = \tilde{\tc{A}}^{-1} \boldsymbol{\hat{f}}_{\phi} = \tilde{\tc{A}}^{-1} \tc{T} \boldsymbol{\hat{f}}.
\end{equation}

Since (\ref{Eq:SemiDiscPDE}) is hyperbolic, the solution $\boldsymbol{\hat{\phi}}_{\pm}$ of (\ref{Eq:EllipticBVP}) consists of a summation of downstream- and upstream-traveling modes, i.e., waves that transfer energy in the positive and negative $x$-direction, respectively (waves that do not propagate were eliminated by the contraction of the system that lead to $\tc{M}$).  These downstream- and upstream-traveling components of the solution of~(\ref{Eq:EllipticBVP}) can be identified based on the eigenvalues and eigenvectors of $\tc{M}$.  Consider the eigen-expansion of the solution
\begin{equation}
\label{Eq:eigen_expansion}
\boldsymbol{\hat{\phi}}_{\pm}(x,s) = \sum_{k = 1}^{N} \boldsymbol{v}_{k}(x,s) \boldsymbol{\psi}_{k}(x,s),
\end{equation}
where each $\boldsymbol{v}_{k}$ is an eigenvector of $\tc{M}$ with an associated eigenvalue $i \alpha_{k}$, and $\boldsymbol{\psi}_{k}$ is an expansion coefficient defining the contribution of mode $k$ to the solution.  The well-posedness theory of \cite{Kreiss:1970}, which can be thought of as an extension to $x$-dependent systems of Briggs' criteria \citep{Briggs:1964}, provides a means to distinguish downstream- and upstream-traveling components of the solution: the mode associated with the eigenvalue $\alpha_{k}(x,s)$ is downstream-traveling at $x = x_{0}$ if
\begin{equation}
\label{Eq:rightgoing_X}
\lim_{\eta \to +\infty} \ImV\left[\alpha_{k}\left(x_{0},s\right)\right] = +\infty
\end{equation}
and upstream-traveling if 
\begin{equation}
\label{Eq:leftgoing_X}
\lim_{\eta \to +\infty} \ImV\left[\alpha_{k}\left(x_{0},s\right)\right] = -\infty.
\end{equation}

The eigenvalue decomposition of $\tc{M}$ can be written
\begin{equation}
\label{Eq:M_eig_decomp}
\tc{M} = \left[ \begin{array}{cc} \tc{V}_{+} & \tc{V}_{-} \end{array} \right] \left[ \begin{array}{cc} \tc{D}_{++} & \tc{0} \\ \tc{0} & \tc{D}_{--} \end{array} \right] \left[ \begin{array}{c} \tc{U}_{+} \\  \tc{U}_{-} \end{array} \right],
\end{equation}
where the columns of $\tc{V}_{+} \in \mathbb{C}^{N \times N_{+}}$ and $\tc{V}_{-}\in \mathbb{C}^{N \times N_{-}}$ and the rows of $\tc{U}_{+}\in \mathbb{C}^{N_{+} \times N}$ and $\tc{U}_{-}\in \mathbb{C}^{N_{-} \times N}$ contain the left and right eigenvectors associated with the downstream- and upstream-traveling eigenvalues of $\tc{M}$, respectively, which are contained in the diagonal matrices $\tc{D}_{++} \in \mathbb{C}^{N_{+} \times N_{+}}$ and $\tc{D}_{--}\in \mathbb{C}^{N_{-} \times N_{-}}$.  The eigenvectors are normalized such that
\begin{equation}
\label{Eq:biorthonormal}
\left[ \begin{array}{cc} \tc{V}_{+} & \tc{V}_{-} \end{array} \right] \left[ \begin{array}{c} \tc{U}_{+} \\  \tc{U}_{-} \end{array} \right] = \left[ \begin{array}{c} \tc{U}_{+} \\  \tc{U}_{-} \end{array} \right] \left[ \begin{array}{cc} \tc{V}_{+} & \tc{V}_{-} \end{array} \right] = \tc{I}.
\end{equation}

Using this block matrix notation, (\ref{Eq:eigen_expansion}) can be written as
\begin{equation}
\label{Eq:solution_expansion2}
\boldsymbol{\hat{\phi}}_{\pm} = \tc{V} \boldsymbol{\psi} = \tc{V}_{+} \boldsymbol{\psi}_{+} + \tc{V}_{-} \boldsymbol{\psi}_{-},
\end{equation}
where 
\begin{equation}
\label{Eq:psi_definition}
\boldsymbol{\psi} = \left\{\begin{array}{c} \boldsymbol{\psi}_{+} \\ \boldsymbol{\psi}_{-}  \end{array}\right\}
\end{equation}
and $\boldsymbol{\psi}_{+}$ and $\boldsymbol{\psi}_{-}$ are vectors of expansion coefficients for the downstream- and upstream-traveling modes, respectively.  Therefore, the downstream-traveling part of the solution is
\begin{equation}
\label{Eq:phi_prime}
\boldsymbol{\hat{\phi}}_{\pm}' = \tc{V}_{+} \boldsymbol{\psi}_{+}
\end{equation}
and the upstream-traveling part is
\begin{equation}
\label{Eq:phi_primeprime}
\boldsymbol{\hat{\phi}}_{\pm}'' = \tc{V}_{-} \boldsymbol{\psi}_{-}.
\end{equation}

\subsection{Exact projection operator}
\label{Sec:ExactProjection}

We define a projection operator 
\begin{equation}
\label{Eq:projection_exact}
\tc{P} = \left[ \begin{array}{cc} \tc{V}_{+} & \tc{V}_{-} \end{array} \right] \left[ \begin{array}{cc} \,\,\,\tc{I}\,\,\, & \tc{0} \\ \tc{0} & \,\,\,\tc{0}\,\,\, \end{array} \right] \left[ \begin{array}{c} \tc{U}_{+} \\  \tc{U}_{-} \end{array} \right]
\end{equation}
that exactly splits the solution $\boldsymbol{\hat{\phi}}_{\pm}$ into downstream- and upstream-traveling components at each $x$.  That is, 
\begin{subequations}
\begin{align}
&\boldsymbol{\hat{\phi}}_{\pm}' = \tc{P} \boldsymbol{\hat{\phi}}_{\pm}, \label{Eq:phi_prime_P} \\
&\boldsymbol{\hat{\phi}}_{\pm}'' = \left(\tc{I} - \tc{P}\right) \boldsymbol{\hat{\phi}}_{\pm}. \label{Eq:phi_primeprime_P}
\end{align}
\end{subequations}
Equation~(\ref{Eq:phi_prime_P}) follows from~(\ref{Eq:solution_expansion2})~and~(\ref{Eq:phi_prime}):
\begin{subequations}
\label{Eq:phiprime_eq_P_phi}
	\begin{align}
	\tc{P} \boldsymbol{\hat{\phi}}_{\pm} &= \left[ \begin{array}{cc} \tc{V}_{+} & \tc{V}_{-} \end{array} \right] \left[ \begin{array}{cc} \,\,\,\tc{I}\,\,\, & \tc{0} \\ \tc{0} & \,\,\,\tc{0}\,\,\, \end{array} \right] \left[ \begin{array}{c} \tc{U}_{+} \\  \tc{U}_{-} \end{array} \right] \left[ \begin{array}{cc} \tc{V}_{+} & \tc{V}_{-} \end{array} \right] \left[ \begin{array}{c} \boldsymbol{\psi}_{+} \\  \boldsymbol{\psi}_{-} \end{array} \right] \\
	&=  \tc{V}_{+} \boldsymbol{\psi}_{+} \\
	&=  \boldsymbol{\hat{\phi}}_{\pm}',
	\end{align}
\end{subequations}
and (\ref{Eq:solution_expansion2})~and~(\ref{Eq:phi_primeprime}) can be similarly manipulated to verify~(\ref{Eq:phi_primeprime_P}).  Using (\ref{Eq:projection_exact}) and (\ref{Eq:biorthonormal}), it is straightforward to show that $\tc{P}$ is a projection operator, i.e., that  $\tc{P} \tc{P} = \tc{P}$.


\subsection{One-way equation}
\label{sec:OWE_projection_onewayEq}

We now obtain a one-way equation for the downstream-traveling component of the solution $\boldsymbol{\hat{\phi}}_{\pm}'$ by applying the projection $\tc{P}$ to~(\ref{Eq:EllipticBVP}).  This gives
\begin{equation}
\label{Eq:EllipticBVP_P}
\tc{P} \frac{d \boldsymbol{\hat{\phi}}_{\pm}}{dx} = \tc{P} \tc{M}\boldsymbol{\hat{\phi}}_{\pm} + \tc{P} \boldsymbol{\hat{h}}.
\end{equation}
To obtain an evolution equation for $\boldsymbol{\hat{\phi}}_{\pm}'$, we must move $\tc{P}$ inside the derivative.  Using the chain rule, we have
\begin{equation}
\label{Eq:P_inside}
\frac{d \boldsymbol{\hat{\phi}}_{\pm}'}{dx} = \frac{d \tc{P} \boldsymbol{\hat{\phi}}_{\pm}}{dx} = \tc{P} \frac{d \boldsymbol{\hat{\phi}}_{\pm}}{dx} + \frac{d \tc{P} }{dx} \boldsymbol{\hat{\phi}}_{\pm}.
\end{equation}
Using that $\tc{P}$ and $\tc{M}$ commute (since they have the same eigenvectors by~(\ref{Eq:M_eig_decomp})~and~(\ref{Eq:projection_exact})), the first term on the right-hand-side of~(\ref{Eq:EllipticBVP_P}) can also be written in terms of $\boldsymbol{\hat{\phi}}_{\pm}'$,
\begin{equation}
\label{Eq:PMphi}
\tc{P}\tc{M}\boldsymbol{\hat{\phi}}_{\pm} = \tc{P} \tc{P}\tc{M}\boldsymbol{\hat{\phi}}_{\pm} =  \tc{P}\tc{M} \tc{P} \boldsymbol{\hat{\phi}}_{\pm} = \tc{P}\tc{M} \boldsymbol{\hat{\phi}}_{\pm}'.
\end{equation}

Therefore, using~(\ref{Eq:P_inside})~and~(\ref{Eq:PMphi}), (\ref{Eq:EllipticBVP_P}) can be written
\begin{equation}
\label{Eq:phi_prime_ODE_exact}
\frac{d \boldsymbol{\hat{\phi}}_{\pm}'}{dx} = \tc{P} \left( \tc{M}\boldsymbol{\hat{\phi}}_{\pm}' + \boldsymbol{\hat{h}}\right) + \frac{d \tc{P} }{dx} \boldsymbol{\hat{\phi}}_{\pm}.
\end{equation}
Following the same steps, but with $\tc{I}-\tc{P}$ replacing $\tc{P}$, we similarly obtain
\begin{equation}
\label{Eq:phi_primeprime_ODE_exact}
\frac{d \boldsymbol{\hat{\phi}}_{\pm}''}{dx} = \left( \tc{I}-\tc{P} \right) \left( \tc{M}\boldsymbol{\hat{\phi}}_{\pm}'' + \boldsymbol{\hat{h}}\right) - \frac{d \tc{P} }{dx} \boldsymbol{\hat{\phi}}_{\pm}.
\end{equation}

By neglecting $\frac{d \tc{P}}{dx} \boldsymbol{\hat{\phi}}_{\pm}$ in~(\ref{Eq:phi_prime_ODE_exact})~and~(\ref{Eq:phi_primeprime_ODE_exact}), we arrive at one-way equations for the downstream- and upstream-traveling components of the solution,
\begin{equation}
\label{Eq:OWE-P1}
\frac{d \boldsymbol{\hat{\phi}}_{\pm}'}{dx} = \tc{P} \left( \tc{M}\boldsymbol{\hat{\phi}}_{\pm}' + \boldsymbol{\hat{h}}\right),
\end{equation}
\begin{equation}
\label{Eq:OWE-P2}
\frac{d \boldsymbol{\hat{\phi}}_{\pm}''}{dx} = \left( \tc{I}-\tc{P} \right) \left( \tc{M}\boldsymbol{\hat{\phi}}_{\pm}'' + \boldsymbol{\hat{h}}\right).
\end{equation}

When $\tc{M}$ is $x$-independent, $\frac{d \tc{P} }{dx} = \boldsymbol{0}$ and~(\ref{Eq:OWE-P1})~and~(\ref{Eq:OWE-P2}) exactly describe the evolution of downstream- and upstream-traveling waves, respectively.  When $\tc{M}$ is $x$-dependent, $\frac{d \tc{P} }{dx} \neq \boldsymbol{0}$ and~(\ref{Eq:OWE-P1})~and~(\ref{Eq:OWE-P2}) are approximate.  Insight into the nature of the approximation can be gained by solving both~(\ref{Eq:phi_prime_ODE_exact}) and~(\ref{Eq:phi_primeprime_ODE_exact}) for $\frac{d \tc{P}}{dx} \boldsymbol{\hat{\phi}}_{\pm}$ and equating the expressions, giving
\begin{equation}
\label{Eq:OWE-P_both}
\frac{d \boldsymbol{\hat{\phi}}_{\pm}'}{dx} - \tc{P} \left( \tc{M}\boldsymbol{\hat{\phi}}_{\pm}' + \boldsymbol{\hat{h}}\right) = -\left( \frac{d \boldsymbol{\hat{\phi}}_{\pm}''}{dx} - \left( \tc{I}-\tc{P} \right) \left( \tc{M}\boldsymbol{\hat{\phi}}_{\pm}'' + \boldsymbol{\hat{h}}\right) \right) =  \frac{d \tc{P}}{dx} \boldsymbol{\hat{\phi}}_{\pm}.
\end{equation}
Comparing~(\ref{Eq:OWE-P1})~and~(\ref{Eq:OWE-P_both}) reveals that neglecting $\frac{d \tc{P}}{dx} \boldsymbol{\hat{\phi}}_{\pm}$ is equivalent to setting $\boldsymbol{\phi}'' = 0$ when calculating $\boldsymbol{\phi}'$.  In other words, the one-way equation~(\ref{Eq:OWE-P1}) neglects the influence of the upstream-traveling waves on the evolution of the downstream-traveling waves.  In the same way, comparing~(\ref{Eq:OWE-P2})~and~(\ref{Eq:OWE-P_both}) reveals that neglecting $\frac{d \tc{P}}{dx} \boldsymbol{\hat{\phi}}_{\pm}$ is equivalent to setting $\boldsymbol{\phi}' = 0$ when calculating $\boldsymbol{\phi}''$; the one-way equation~(\ref{Eq:OWE-P2}) neglects the influence of the downstream-traveling waves on the evolution of the upstream-traveling waves.  As discussed in detail by \cite{Towne2015oneway}, it is reasonable to neglect the influence of upstream-traveling waves on the downstream-traveling waves (and vice versa) when $\tc{M}$ is slowly varying in $x$.  Since $\tc{M}$ inherits its $x$-dependence from the mean flow $\bar{q}$, the one-way equation will yield an accurate approximation of the downstream- or upstream-traveling response to the forcing for slowly varying flows.  

Next, we will show that~(\ref{Eq:OWE-P1})~and~(\ref{Eq:OWE-P2}) are well-posed as one-way equations.  This amounts to showing that their eigenvalues correspond to downstream- and upstream-traveling modes, respectively.  Focusing first on~(\ref{Eq:OWE-P1}), the relevant operator is 
\begin{subequations}
\label{Eq:PM}
	\begin{align}
	\tc{P}\tc{M} &= \left[ \begin{array}{cc} \tc{V}_{+} & \tc{V}_{-} \end{array} \right] \left[ \begin{array}{cc} \,\,\,\tc{I}\,\,\, & \tc{0} \\ \tc{0} & \,\,\,\tc{0}\,\,\, \end{array} \right] \left[ \begin{array}{c} \tc{U}_{+} \\  \tc{U}_{-} \end{array} \right] \left[ \begin{array}{cc} \tc{V}_{+} & \tc{V}_{-} \end{array} \right] \left[ \begin{array}{cc} \tc{D}_{++} & \tc{0} \\ \tc{0} & \tc{D}_{--} \end{array} \right] \left[ \begin{array}{c} \tc{U}_{+} \\  \tc{U}_{-} \end{array} \right] \\
	&= \left[ \begin{array}{cc} \tc{V}_{+} & \tc{V}_{-} \end{array} \right] \left[ \begin{array}{cc} \tc{D}_{++} & \tc{0} \\ \tc{0} & \,\,\,\tc{0}\,\,\, \end{array} \right]  \left[ \begin{array}{c} \tc{U}_{+} \\  \tc{U}_{-} \end{array} \right].
	\end{align}
\end{subequations}
Compared to the original elliptic operator $\tc{M}$, the eigenvectors and downstream-traveling eigenvalues of the one-way operator $\tc{P}\tc{M}$ are unchanged, but the upstream-traveling eigenvalues have been eliminated.  Therefore, (\ref{Eq:OWE-P1}) is well-posed as a one-way equation and can be solved by integrating the equations in the positive $x$ direction.

The relevant operator for the well-posedness of~(\ref{Eq:OWE-P2}) is
\begin{subequations}
\label{Eq:M-PM}
	\begin{align}
	\left(\tc{I}-\tc{P}\right) \tc{M} &= \left[ \begin{array}{cc} \tc{V}_{+} & \tc{V}_{-} \end{array} \right] \left[ \begin{array}{cc} \,\,\,\tc{0}\,\,\, & \tc{0} \\ \tc{0} & \,\,\,\tc{I}\,\,\, \end{array} \right] \left[ \begin{array}{c} \tc{U}_{+} \\  \tc{U}_{-} \end{array} \right] \left[ \begin{array}{cc} \tc{V}_{+} & \tc{V}_{-} \end{array} \right] \left[ \begin{array}{cc} \tc{D}_{++} & \tc{0} \\ \tc{0} & \tc{D}_{--} \end{array} \right] \left[ \begin{array}{c} \tc{U}_{+} \\  \tc{U}_{-} \end{array} \right] \\
	&= \left[ \begin{array}{cc} \tc{V}_{+} & \tc{V}_{-} \end{array} \right] \left[ \begin{array}{cc} \,\,\,\tc{0}\,\,\, & \tc{0} \\ \tc{0} & \tc{D}_{--} \end{array} \right]  \left[ \begin{array}{c} \tc{U}_{+} \\  \tc{U}_{-} \end{array} \right].
	\end{align}
\end{subequations}
The downstream-traveling eigenvalues have been eliminated, so~(\ref{Eq:OWE-P2}) is well-posed as a one-way equation and can be solved by integrating the equations in the negative $x$ direction.


This projection-based paradigm for obtaining one-way equations and the projection operator defined in~(\ref{Eq:projection_exact}) were originally derived by \cite{Towne:2016} and was recently rederived by \cite{Harris2020well}.  While it is well posed, as we have shown, its computational efficiency is problematic; the eigen-decomposition of $\tc{M}$ is required at every $x$ to construct the exact projection $\tc{P}$, resulting in an intolerably high computational cost for large $N$ due to the nominal $O(N^{3})$ scaling of the number of operations required to solve each eigenvalue problem.  To obtain a practically useful one-way equation, we construct in the next section an approximation of $\tc{P}$ that can be efficiently computed.


\subsection{Approximate projection operator}
\label{sec:OWE_projection_approxP}


The following set of recursion equations approximate the action of $\tc{P}$ on an arbitrary vector $\boldsymbol{\hat{\phi}}_{\pm}$:
\begin{subequations}
\label{Eq:OWEP_recursions}
	\begin{align}
	\label{Eq:OWEP_recursions_a} &\,\boldsymbol{\hat{\phi}}^{-N_{\beta}}_{+} = 0  \\
	\label{Eq:OWEP_recursions_b} &\left(\tc{M} - i \beta_{-}^{j} \tc{I} \right) \boldsymbol{\hat{\phi}}_{\pm}^{-j} - \left(\tc{M} - i \beta_{+}^{j} \tc{I} \right) \boldsymbol{\hat{\phi}}_{\pm}^{-j-1} = 0 \qquad \qquad  &&j = 1,\dots,N_{\beta}-1   \\
	\label{Eq:OWEP_recursions_c}&\left(\tc{M} - i \beta_{-}^{0} \tc{I} \right) \boldsymbol{\hat{\phi}}_{\pm}^{0} - \left(\tc{M} - i \beta_{+}^{0} \tc{I} \right) \boldsymbol{\hat{\phi}}_{\pm}^{-1} = \left(\tc{M} - i  \beta_{-}^{0} \tc{I} \right) \boldsymbol{\hat{\phi}}_{\pm}  \\
	\label{Eq:OWEP_recursions_d}&\left(\tc{M} - i \beta_{+}^{j} \tc{I} \right) \boldsymbol{\hat{\phi}}_{\pm}^{j} - \left(\tc{M} - i \beta_{-}^{j} \tc{I} \right) \boldsymbol{\hat{\phi}}_{\pm}^{j+1} = 0 \qquad \qquad  &&j = 0,\dots,N_{\beta}-1  \\
	\label{Eq:OWEP_recursions_e}&\,\boldsymbol{\hat{\phi}}^{N_{\beta}}_{-} = 0. 
	\end{align}
\end{subequations}
Here, we have introduced a set of auxiliary variables $\{ \boldsymbol{\hat{\phi}}_{\pm}^{j}: j = -N_{\beta},\dots,N_{\beta} \}$ and a set of complex scalar recursion parameters $\{ \beta^{\,j}_{+}, \beta^{\,j}_{-}: j = 0,\dots,N_{\beta}-1 \}$.  $N_{\beta}$ is the order of the approximate projection.  

In Appendix~\ref{App:Recursions}, we show that the zero-indexed variable is the approximate projection of $\boldsymbol{\hat{\phi}}_{\pm}$, i.e., that
\begin{equation}
\label{Eq:ApproxPhiPrime}
\boldsymbol{\hat{\phi}}_{\pm}^{0} = \tc{P}_{N_{\beta}} \boldsymbol{\hat{\phi}}_{\pm} \approx \boldsymbol{\hat{\phi}}_{\pm}'.
\end{equation}

The operator 
\begin{equation}
\label{Eq:OWEP_P_approx}
\tc{P}_{N_{\beta}} = \left[ \begin{array}{cc} \tc{V}_{+} & \tc{V}_{-} \end{array} \right] \left[\begin{array}{cc} \tc{I} & -\tc{R}_{+-} \\ -\tc{R}_{-+} & \tc{I} \end{array}\right]^{-1} \left[ \begin{array}{cc} \,\,\,\tc{I}\,\,\, & \tc{0} \\ \tc{0} & \,\,\,\tc{0}\,\,\, \end{array} \right] \left[\begin{array}{cc} \tc{I} & -\tc{R}_{+-} \\ -\tc{R}_{-+} & \tc{I} \end{array}\right] \left[ \begin{array}{c} \tc{U}_{+} \\  \tc{U}_{-} \end{array} \right]
\end{equation}
is the approximation of the exact projection $\tc{P}$ that is implicitly defined by the recursions~(\ref{Eq:OWEP_recursions}).  It is easy to verify that $\tc{P}_{N_{\beta}} \tc{P}_{N_{\beta}} = \tc{P}_{N_{\beta}}$, so $\tc{P}_{N_{\beta}}$ is itself a projection.  Furthermore, comparing~(\ref{Eq:projection_exact})~and~(\ref{Eq:OWEP_P_approx}), we see that $\tc{P}_{N_{\beta}} \to \tc{P}$ as $\tc{R}_{+-},\tc{R}_{+-} \to \tc{0}$.  Therefore, the approximation converges if every entry of $\tc{R}_{+-}$ and $\tc{R}_{+-}$ converge toward zero as the order of the approximation increases.  In appendix~\ref{App:Recursions}, we show that the $(n,m)$ entry of  $\tc{R}_{+-}$ and the $(m,n)$ entry of $\tc{R}_{-+}$ are, respectively, 
\begin{subequations}
\label{Eq:OWEP_R_entries}
\begin{align}
&\left(\tc{R}_{+-}\right)_{nm} = \frac{\mathcal{F}(\alpha_{+,n})}{\mathcal{F}(\alpha_{-,m})} \left(\boldsymbol{w}_{+-}\right)_{nm}, \\
&\left(\tc{R}_{-+}\right)_{mn} = \frac{\mathcal{F}(\alpha_{+,n})}{\mathcal{F}(\alpha_{-,m})} \left(\boldsymbol{w}_{-+}\right)_{mn},
\end{align}
\end{subequations}
where 
\begin{equation}
\label{Eq:ApproxParDAErecursionR}
\mathcal{F}(\alpha) = \prod_{j=0}^{N_{\beta}-1} \frac{\alpha-\beta^{\,j}_{+}}{\alpha-\beta^{\,j}_{-}},
\end{equation}
$\alpha_{+,n}$ is the $n^{\text{th}}$ downstream-traveling eigenvalue, $\alpha_{-,m}$ is the $m^{\text{th}}$ upstream-traveling eigenvalue, and $\left(\boldsymbol{w}_{+-}\right)_{nm}$ and $\left(\boldsymbol{w}_{-+}\right)_{mn}$ are scalar weights that do not depend on the recursion parameters.  Since the weights are fixed, the recursion parameters must be chosen such that $\mathcal{F}(\alpha_{+,n})/\mathcal{F}(\alpha_{-,m})$ goes to zero for all $m,n$.  

A geometric interpretation of $\mathcal{F}$ is helpful for choosing parameters that accomplish this objective.  Notice that the magnitude of each term in the product defining $\mathcal{F}$ is less than one for regions of the complex $\alpha$ plane that are closer to $\beta_{+}^{j}$ than $\beta_{-}^{j}$ ($|\alpha - \beta_{+}^{j}| < |\alpha - \beta_{-}^{j}|$) and greater than one for regions that are closer to $\beta_{-}^{j}$ than $\beta_{+}^{j}$ ($|\alpha - \beta_{+}^{j}| > |\alpha - \beta_{-}^{j}|$).  Therefore, $\mathcal{F}(\alpha_{+,n})$ is driven to zero by placing the $\beta_{+}^{j}$ parameters near the downstream-traveling eigenvalues in the complex plane, and $\mathcal{F}(\alpha_{-,m})$ is driven to infinity by placing the $\beta_{-}^{j}$ parameters near the upstream-traveling eigenvalues.


This is exactly the same requirement for convergence as derived for the OWNS-O method by \cite{Towne2015oneway}.  This has several important implications.  First, as established by \cite{Towne2015oneway}, if $\alpha_{+,n} \neq \alpha_{-,m}$ for all $m$,$n$, there always exist recursion parameters that make the approximation error arbitrarily small.  Second, if the recursion parameters are well-placed, the convergence of the approximation is exponential.  Third, any recursion parameters derived for OWNS-O can be used without modification for OWNS-P.  A general strategy for choosing recursion parameters was outlined by \cite{Towne2015oneway}, and effective sets of recursion parameters have been developed for mixing layers \citep{Towne:2014}, jets \citep{Towne:2013}, subsonic boundary layers \citep{Rigas:2017a}, and supersonic boundary layers \citep{Kamal2020application}.

Finally, to be rigorous, we must show that~(\ref{Eq:OWE-P1})~and~(\ref{Eq:OWE-P2}) remain well-posed as one-way equations when $\tc{P}_{N_{\beta}}$ is used in place of $\tc{P}$.  This is confirmed in Appendix~\ref{App:WellPosed}.

\subsection{Implementation}
\label{sec:implemenation}

The recursion equations defining the approximate projection operator define a system of equations of the form
\begin{subequations}
\label{Eq:OWEP_recursions_system}
\begin{align}
\boldsymbol{\hat{\phi}}_{\pm}' = \tc{P}_3 \boldsymbol{\hat{\phi}}^{\mathrm{aux}} \label{Eq:OWEP_recursions_system_a} \\
\tc{P}_2 \boldsymbol{\hat{\phi}}^{\mathrm{aux}} = \tc{P}_1 \boldsymbol{\hat{\phi}}_{\pm} \label{Eq:OWEP_recursions_system_b}
\end{align}
\end{subequations}
where $\boldsymbol{\hat{\phi}}^{\mathrm{aux}} \in \mathbb{C}^{N_{\mathrm{aux}}}$ is a vector containing all of the auxiliary variables, the matrices $\tc{P}_{1} \in \mathbb{C}^{N_{\mathrm{aux}} \times N}$ and $\tc{P}_{2} \in \mathbb{C}^{N_{\mathrm{aux}} \times N_{\mathrm{aux}}}$ are defined by the recursion equations~(\ref{Eq:OWEP_recursions}), $\tc{P}_{3} \in \mathbb{C}^{N \times N_{\mathrm{aux}}}$ is a matrix that extracts the projected state from the auxiliary variables via~(\ref{Eq:ApproxPhiPrime}), and $N_{\mathrm{aux}} = 2 N N_{\beta} + N_{0}$.  The structure of these matrices is exemplified in Appendix~\ref{App:OWNS_mats}.

From~(\ref{Eq:OWEP_recursions_system}) we see that applying the projection operator to a vector $\boldsymbol{\hat{\phi}}_{\pm}$ to obtain the projected state $\boldsymbol{\hat{\phi}}_{\pm}'$ requires the solution of an a linear system of size $N_{\mathrm{aux}}$; the cost of this operation compared to those associated with a global solution strategy is discussed in the next section.  We stress that in practice we never form the approximate projection operator $\tc{P}_{N_{\beta}}$.  

The approximate form of the one-way equation~(\ref{Eq:OWE-P1}) can be expressed as a DAE input/output system 
\begin{subequations}
\label{Eq:DAE_OWNSP_approx}
\begin{align}
\tc{A}^\ddagger \frac{d \boldsymbol{\hat{\phi}}^\ddagger}{dx} &= \tc{L}^\ddagger \boldsymbol{\hat{\phi}}^\ddagger + \tc{B}^{\ddagger} \boldsymbol{\hat{f}}_{\phi}, \label{Eq:DAE_OWNSP_approx_a} \\
\boldsymbol{\hat{\phi}}' &= \tc{C}^\ddagger \boldsymbol{\hat{\phi}}^\ddagger. \label{Eq:DAE_OWNSP_approx_b}
\end{align}
\end{subequations}
The expanded state vector is
\begin{equation}\label{Eq:DAE_OWNSP_state}
\boldsymbol{\phi}^\ddagger = \begin{bmatrix} \boldsymbol{\hat{\phi}}_{\pm}' \\ \boldsymbol{\hat{\phi}}_{0}' \\ \boldsymbol{\hat{\phi}}^{\mathrm{aux}} \end{bmatrix}.
\end{equation}
and the operators in~(\ref{Eq:DAE_OWNSP_approx}) are
\begin{equation}\label{eq:approx_POWNS}
\tc{A}^\ddagger = \begin{bmatrix} \tc{I} &  &  \\  & \tc{0} &  \\  &  & \tc{0}  \end{bmatrix}, \quad \tc{B}^\ddagger = \begin{bmatrix} \tc{0} & \tc{0} \\ \tc{P}_1\tilde{\tc{A}}_{\pm\pm}^{-1} & \tc{0}  \\ \tc{0} & \tc{I}  \end{bmatrix}, \quad \tc{C}^{\ddagger} = \begin{bmatrix} \tc{I} & \tc{0} & \tc{0} \\ \tc{0} & \tc{I} & \tc{0} \end{bmatrix},
\end{equation}
and
\begin{equation}\label{eq:approx_POWNS_L}
\tc{L}^\ddagger = \begin{bmatrix} \tc{0} & \tc{0} & \tc{P}_3\\ \tc{P}_1\tilde{\tc{A}}_{\pm\pm}^{-1} \tc{L}_{\pm\pm} & \tc{P}_1\tilde{\tc{A}}_{\pm\pm}^{-1} \tc{L}_{\pm0} & -\tc{P}_2 \\ \tc{L}_{0\pm} & \tc{L}_{00} & \tc{0} \end{bmatrix}.
\end{equation}
The input to the system is the forcing $\boldsymbol{\hat{f}}_{\phi} = \tc{T} \boldsymbol{\hat{f}}$, while the output is $\boldsymbol{\hat{\phi}}'$, the downstream-traveling component of the characteristic variable, from which the OWNS-P approximation of the physical state variable can be obtained using~(\ref{Eq:CharVars}) as $\boldsymbol{\hat{q}}' = \tc{T}^{-1} \boldsymbol{\hat{\phi}}'$.  The DAE~(\ref{Eq:DAE_OWNSP_approx}) can be efficiently integrated in the positive $x$-direction given a specified value for $\boldsymbol{\hat{\phi}}_{\pm}'$ at the inlet of the domain; this value is physically a boundary condition for the global problem but functions as an initial condition for the spatial integration of the DAE.  


An additional issue arises in the practical implementation of the method.  Specifically, errors incurred during the numerical integration of~(\ref{Eq:DAE_OWNSP_approx}) will not lie entirely in the downstream-traveling subspace.  In other words, the numerical approximation of $\boldsymbol{\hat{\phi}}_{\pm}'$ collects an error that projects onto the zero eigenvalues of $\tc{P}_{N_{\beta}} \tc{M}$, which is then propagated along during the march.  This causes an accumulation of error that contaminates the solution.  Fortunately, there is an easy fix: apply the projection operator to the solution after each step in the march.  


\subsection{Computational cost}
\label{sec:OWE_projection_cost}

The standard approach for obtaining the action of the resolvent operator on a forcing vector requires the solution of a linear system of equations of size $N_{q} N_{x} N_{c}$, where $N_{x}$ and $N_{c}$ are the number of discretizations points in $x$ and in all transverse directions, respectively, and $N_{q}$ is the number of state variables, e.g., $N_{q} = 5$.  In the following scaling estimates, we drop the dependence on $N_{q}$ since it is a constant for a given problem.  Assuming the use of sparse discretization schemes, and direct solution via multi-frontal LU decomposition, the CPU cost (FLOPS) of solving the linear system is found, empirically, to scale as $O(N_{x}^{a} N_{c}^{a})$, and the memory usage scales as  $O(N_{x}^{b} N_{c}^{b})$, where the factors $1 < a \le 3$ and $1 < b \le 2$ depend on the sparsity and structure of the matrix and the sophistication of the algorithm employed \citep{Duff2017direct}.  In our global computations for 2D base flows corresponding to turbulent jets presented in part 2 \citep{Rigas2020fast}, we observed $a \approx 1.6$ and $b \approx 1.2$, whereas in some preliminary computations for 3D base flows, we observed $a \approx 2$ and $b \approx 2$.  We have also implemented iterative solvers based on GMRES and Bi-CG-Stab, which decrease both $a$ and $b$ to near unity, but, without preconditioning, these were typically slower as the number of iterations required grew large.

The main cost for OWNS-P is solving the system of equations~(\ref{Eq:OWEP_recursions_system}) of size $N_{\mathrm{aux}} = 2 N_{\beta}  N + N_{0} = 2 N_{\beta} N_{q} N_{c}  + N_{0}$ if the DAE~(\ref{Eq:DAE_OWNSP_approx}) is explicitly integrated or a system of size $N^{\ddagger} = (2 N_{\beta}+1) N + N_0 = (2 N_{\beta}+1) N_{q} N_{c} + N_0 $ if it is implicitly integrated.  In both cases, the dominant factor in these size expressions is $N_{c} N_{\beta}$, and other terms are dropped.  As the sparsity and structure of the OWNS matrices are analogous to those in the global solution, the FLOPS and memory of solving~(\ref{Eq:OWEP_recursions_system}) scale as $O(N_{x} N_{c}^{a} N_{\beta}^{a})$ and $O(N_{c}^{b} N_{\beta}^{b})$.  The additional factor of $N_{x}$ in the FLOPS scaling follows from the fact that an equation of this form must be solved at each step in the spatial march.  Assuming $N_x > N_\beta$, OWNS-P represents a FLOPS and memory speedup by factors of $N_x^{a-1}/N_\beta^{a}$ and $N_x^{b}/N_\beta^{b}$, respectively.

Clearly OWNS-P will achieve significant savings in memory, and, for sufficiently large $N_x$, significant saving in FLOPS, compared to global methods.  These reductions allow problems that would otherwise require high-performance computing resources to be solved on a laptop. The advantage grows for 3D base flows where the factors $a$ and $b$ are larger.  

Finally, comparing OWNS-P to OWNS-O, the set of recursion equations that must be solved within the OWNS-P method is roughly twice as large as those required for the OWNS-O method; the index of the auxiliary variables for an approximation of order $N_{\beta}$ span the range $\left[ -N_{\beta}, N_{\beta} \right]$ and $\left[ 0, N_{\beta} \right]$ for OWNS-P and OWNS-O, respectively.  This is the price that must be paid to accommodate the inhomogeneous forcing term.

\section{Examples}
\label{Sec:Examples}

In this section, we present two example problems to validate and demonstrate the OWNS-P method.  The starting point for both problems is the compressible Navier-Stokes equations, written here in terms of specific volume $\varv$, the velocity vector $\bf{u}$, and pressure $p$,

\begin{subequations}
\label{Eq:NS_eqs}
\begin{equation}
\label{Eq:continuity}
\frac{D \varv}{Dt}  - \varv \left( \nabla \cdot \bf{u}  \right) = 0,
\end{equation}
\begin{equation}
\label{Eq:mom}
\frac{D \bf{u}}{D t}  + \varv \nabla p = \frac{1}{Re} \varv  \nabla^{2} \bf{u},
\end{equation}
\begin{equation}
\label{Eq:energy}
\frac{D p}{D t}+\gamma p \left( \nabla \cdot \bf{u}  \right) = \frac{\gamma}{Re Pr} \left( \varv \nabla^{2} p + p \nabla^{2} \varv  \right).
\end{equation}
\end{subequations}
All variables have been appropriately non-dimensionalized by an ambient sound speed and density and a problem dependent length-scale.  The fluid is approximated as a perfect gas with specific heat ratio $\gamma$ and constant Reynolds number $Re$ and Prandtl number $Pr$.  We have neglected viscous energy dissipation and assumed that the gradient of the dilatation is small.  As an example, the linear operators in~(\ref{Eq:LinearPDE_f}) obtained from this form of the Navier-Stokes equations using two-dimensional Cartesian coordinates are reported in Appendix~\ref{App:NS_ops}.
 
For both example problems, the linearized equations are discretized in inhomegeneous transverse directions using fourth-order central finite differences with summation-by-parts boundary closure \citep{Strand1994summation, Mattsson:2004}. Far-field radiation boundary conditions are enforced at free transverse boundaries using a super-grid damping layer \citep{Appelo:2009} truncated by Thompson characteristic conditions \citep{Thompson:1987}. The numerical treatment of the $x$-direction is slightly different in each problem and is reported in the following subsections.  Recursion parameters are selected using the strategies described in \cite{Towne2015oneway}.


\subsection{Dipole forcing of a quiescent fluid}
\label{sec:OWE_projection_validation_dipole}

\begin{figure}
\input{./Figures/Dipole_ExactAndF_2020.tex}
\centering
\includegraphics[trim=0cm 0.0cm 0cm 0.0cm, clip=true]{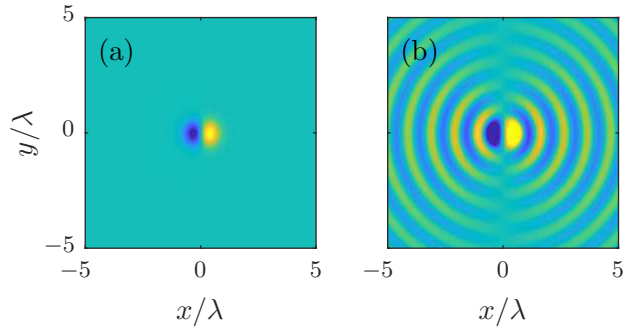}
\caption[Dipole test case: forcing and exact response.]{Dipole test case: (a) forcing applied to the energy equation, and (b) exact pressure response.}
\label{fig:dipole_ExactAndF}
\end{figure}

In this problem, a two-dimensional dipole force is used to excite waves in an inviscid quiescent fluid.  The right-hand-side force that is applied to the energy equation is shown in Figure~\ref{fig:dipole_ExactAndF}(a), and the proper forcing terms are applied to the other equations in order to produce a dipole response in the pressure field, which is shown in Figure~\ref{fig:dipole_ExactAndF}(b).  This is an exact solution of the Euler equations.  

We use the OWNS-P method to obtain an approximate solution.   The computational domain extends ten wavelengths in both $x$ and $y$ and is discretized using $200$ equally spaced points in each direction.  No incoming fluctuations are specified at the domain boundaries -- the waves are excited exclusively by the inhomogeneous forcing terms.

\begin{figure}
\input{./Figures/Dipole_OWE_2020.tex}
\centering
\includegraphics[trim=0cm 0.0cm 0cm 0.0cm, clip=true]{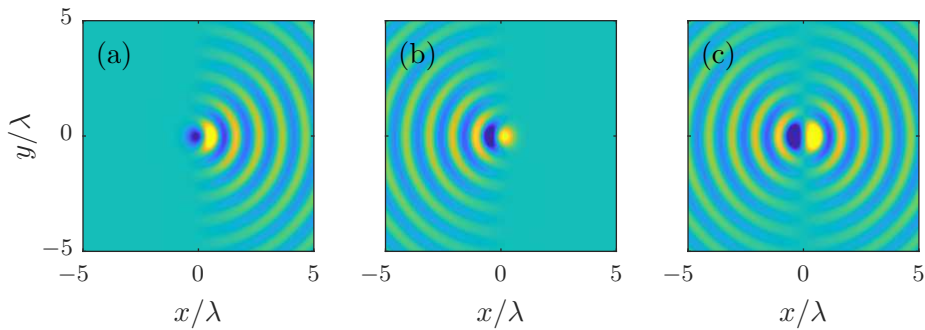}
\caption[One-way approximation of a dipole in a quiescent fluid.]{One-way approximation of the dipole response: (a) downstream-traveling solution, (b) upstream-traveling solution, (c) full solution recovered by summing the downstream- and upstream-traveling solutions.}
\label{fig:dipole_OWE}
\end{figure}

The pressure-field obtained by integrating the one-way equations from left to right is shown in Figure~\ref{fig:dipole_OWE}(a).  Clearly, the downstream-traveling waves are accurately captured.  Similarly, the pressure-field obtained by integrating from right to left is shown in Figure~\ref{fig:dipole_OWE}(b).  This time, the upstream-traveling waves are captured.  Since the governing equations are $x$-independent in this problem, the full solution can be recovered by summing the downstream- and upstream-traveling solutions, which is shown in Figure~\ref{fig:dipole_OWE}(c).

\subsection{Supersonic turbulent jet}
\label{sec:OWE_projection_validation_jet}


Next, we demonstrate our method using the example of a turbulent jet.  Resolvent analysis has been applied to jets by numerous authors \citep{Garnaud:2013, Jeun:2016, Schmidt2018spectral, Lesshafft2019resolvent}, and resolvent modes have been shown to capture a rich set of physical phenomena.  These include the Kelvin-Helmholtz instability, which leads to large-scale coherent wavepacket structures \citep{Jordan:2013}, the Orr mechanism \citep{Tissot2017sensitivity, Schmidt2018spectral}, the lift-up mechanism \citep{Nogueira2019large, Pickering2020lift}, and trapped acoustic waves within the jet core \citep{Tam1989three, Towne2017acoustic}.  The slow spread of the jet makes OWNS applicable, and the diverse set of physics embedded within the resolvent operator make it a challenging test case.

We will make comparisons between a standard global solution of the linearized Navier-Stokes equation (which we will call the LNS solution from here on out) and those obtained using OWNS-P (and OWNS-O and PSE when applicable).  The LNS solution comprises the result of applying the the resolvent operator to a given forcing vector, giving a point of comparison for the approximation of this action provided by the OWNS-P method.

We consider the specific case of a jet with Mach number $M = U_{j} / c_{\infty} = 1.5$, Reynolds number $Re = \rho_{j} U_{j} D / \mu = 1760000$, and temperature ratio $T_{j}/T_{\infty} = 1$, where the subscripts $j$ and $\infty$ denote conditions at the jet nozzle exit and in the far-field, respectively.  The mean flow about which the Navier-Stokes equations are linearized is obtained from a large-eddy simulation described by \cite{Bres2017unstructured}.  Within the linearized equation, we use a turbulent Reynolds number of $Re_T = 1760$, three orders of magnitude less than the true Reynolds number. This choice is motivated by recent work showing that using an eddy-viscosity model or reduced effective Reynolds number improves both the near-field \citep{Pickering2021optimal} and far-field \citep{Pickering2021resolvent} predictions in free-shear flows.

The linearized equations in cylindrical coordinates are discretized in the radial direction as described earlier, and boundary conditions at the polar axis are enforced using the approach of \cite{Mohseni:2002}.  The physical portion of the domain extends to $r/D = 6$ and is discretized using 150 grid points with higher concentration around the shear layer at $x/D = 0.5$, while the damping layer contains an additional 50 points.  Since the mean flow is axisymmetric, the azimuthal direction is homogeneous and can be decomposed into a Fourier series.  In what follows, we focus on the axisymmetric mode, which is often of foremost interest in the study of jet aeroacoustics \citep{Cavalieri2012axisymmetric}.  

The streamwise grid for each method extends from $x/D = 0.5$ to 30 with equidistant spacing of $\Delta x = 0.05$ (with exception to the PSE method, which determines its own step size) and an additional sponge region included in the LNS computation at three diameters upstream and downstream.  The OWNS-O and OWNS-P equations are integrated in the streamwise direction using the 2nd-order backward difference formula and a 5th Order Runge-Kutta Radau II scheme, respectively \citep{Hairer:1991}.  $N_{\beta} = 13$ recursion parameters are used for estimating the OWNS operators. In the following subsections, we show results for two frequencies, $St = 0.26$ and 0.52, which are close to the frequency of maximum acoustic radiation \citep{Jordan:2013} and maximum near-field resolvent gain \citep{Schmidt2018spectral}, respectively.

\subsubsection{Near-nozzle Kelvin-Helmholtz forcing}

First, we compute the downstream response of the linearized equations to a disturbance near the nozzle exit, and compare the OWNS-P solution to those obtained using PSE, OWNS-O, and LNS.  Specifically, the local Kelvin-Helmholtz eigenfunction is specified at $x/D = 0.5$ as an initial condition in the PSE and OWNS marches and as a boundary condition for LNS.  The solution can be interpreted as the response to a localized forcing at the upstream boundary.  In other words, we are computing the result of applying the resolvent operator, approximated by each method, to a forcing vector that is localized at the upstream boundary and which specifically excites the Kelvin-Helmholtz instability.  

\begin{figure}
\centering
\input{./Figures/Jet_compareMethodsIC.tex}
\includegraphics[trim={0cm 0cm 0cm 0.0cm},clip]{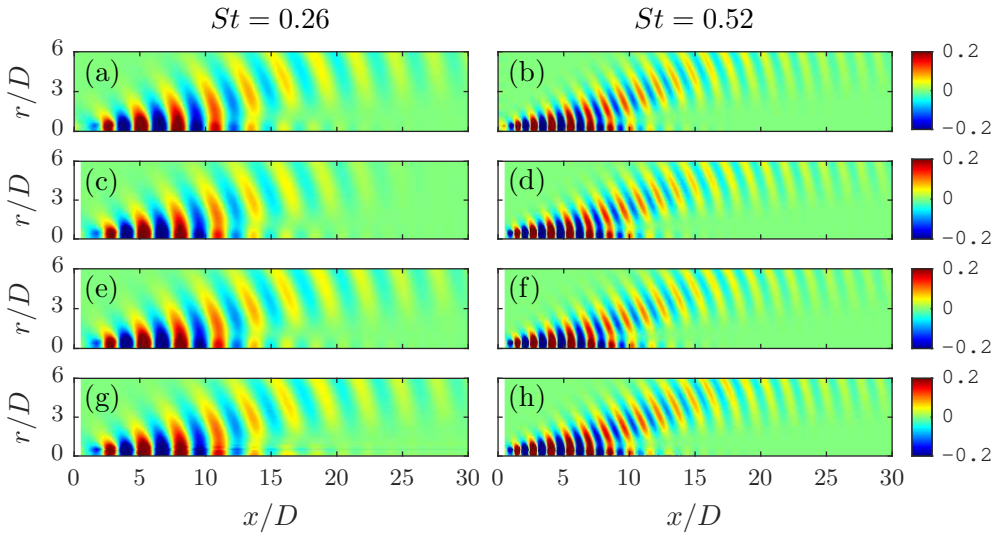}
\caption{Solutions to the Kelvin-Helmholtz initial condition at $x/D = 0.5$ using LNS, PSE, OWNS-O, and OWNS-P methods, top to bottom respectively, for $St = 0.26$, $m=0$, $M=1.5$.  Contours of pressure are shown in each panel.}
\label{fig:ICs}
\end{figure}

Results of this test case are shown in figure~\ref{fig:ICs}, which shows the pressure field computed by each method for $St = 0.26$ and 0.52.  The OWNS-P solution closely matches the LNS solution for both frequencies, demonstrating that the OWNS-P method provides a good approximation of the action of the resolvent operator on the Kelvin-Helmholtz forcing vector.   The OWNS-P solution also matches the OWNS-O solution, which has been previously validated for a similar turbulent jet \citep{Towne2015oneway}.  As shown by \cite{Sinha:2014} and others, PSE also provides a reasonable solution for supersonic jets forced by a Kelvin-Helmholtz mode.  PSE performs well in this case because its underlying assumptions are valid; we have artificially excited a single instability mode by using the Kelvin-Helmholtz mode as an initial condition.  While the acoustic radiation is slightly damped compared to the other three solutions, espeically for $St = 0.26$, it is also reasonably captured because its wavelength is nearly the same as the Kelvin-Helmholtz mode for low-supersonic jets \citep{Towne2019critical}.  However, PSE cannot handle more complex forcings such as those considered next.


\subsubsection{Volumetric forcing using LES data}

Second, we use both OWNS-P and the standard LNS approach to compute the response of the linearized equations to global forcing vectors extracted from LES data.  The right-hand-side forcing term $\boldsymbol{f}$ in~(\ref{Eq:LinearPDE_L}) is obtained by computing the two left-hand-side terms using the LES approximation of the nonlinear Navier-Stokes operator $\mathcal{N}$.  The first term can be computed as
\begin{equation}
\label{Eq:dudt_G}
\frac{\partial q'}{\partial t} = \frac{\partial q}{\partial t} = \mathcal{N}\left(q\right)
\end{equation}
since $\bar{q}$ does not depend on time.  The second term can be approximated as
\begin{equation}
\label{Eq:A_G}
\mathcal{A}\left(\bar{q}\right) q' = \frac{\partial \mathcal{N}}{\partial q}\left(\bar{q}\right)q' \approx \frac{\mathcal{N}\left(\bar{q}+\epsilon q'\right) - \mathcal{N}\left(\bar{q}\right)}{\epsilon}
\end{equation}
for some $\epsilon \ll 1$ \citep{DePando2012efficient}.  We take $\epsilon = 10^{-7}$ and note that the results are nearly independent of $\epsilon$ over a range spanning at least four orders of magnitude.  Details of the procedure can be found in \cite{Towne:2016}.

After performing an azimuthal Fourier transform, an ensemble of realizations of $\boldsymbol{\hat{q}}$ and $\boldsymbol{\hat{f}}$ are obtained from the LES data by segmenting $\boldsymbol{q}^{\prime}$ and $\boldsymbol{f}$ info overlaping blocks, each containing $N_{f}$ snapshots of data.  We use blocks of length $N_{f} = 256$ with 80\% overlap and employ the $w_{C_{4}^{\infty}}$ window proposed by \cite{martini2019accurate} to minimize spectral leakage.  This results in 189 realizations of the flow at discrete frequencies, separated by $\Delta St = 0.02604$. 

Each of these $\boldsymbol{\hat{q}}$, $\boldsymbol{\hat{f}}$ pairs approximately satisfy~(\ref{Eq:q_eq_Rf}).  That is, applying the resolvent operator to $\boldsymbol{\hat{f}}$ yields $\boldsymbol{\hat{q}}$.  This relationship is approximate for the LES data for two reasons.  First, by necessity, we used a finite length DFT in place of an infinite and continuous Fourier transform to obtain $\boldsymbol{\hat{q}}$ and $\boldsymbol{\hat{f}}$, leading to aliasing and spectral leakage.  In particular, $\boldsymbol{\hat{q}}$ exhibits a relatively flat spectrum up to moderate frequencies, making it especially susceptible to aliasing \citep{Towne2017statistical}.  Second, the forcing term $\boldsymbol{f}$ was defined implicitly using the LES operator as described above, whereas our resolvent operator is obtained by explicitly linearizing the Navier-Stokes equations and discretizing the linear equations using numerics different from those within the LES.  Thus, we do not expect our LNS and OWNS-P approximation of the action of the resolvent operator on $\boldsymbol{\hat{f}}$ to exactly reproduce the corresponding $\boldsymbol{\hat{q}}$ obtained from the LES data.  Nevertheless, the $\boldsymbol{\hat{f}}$ vectors obtained from the LES data provide a physically motivated forcing that can be used to compare the action of the resolvent operators obtained via the OWNS-P march and the standard LNS solution.


\begin{figure}
\centering
\input{./Figures/Jet_realizations.tex}
\includegraphics[trim={0cm 0cm 0cm 0.5cm},clip]{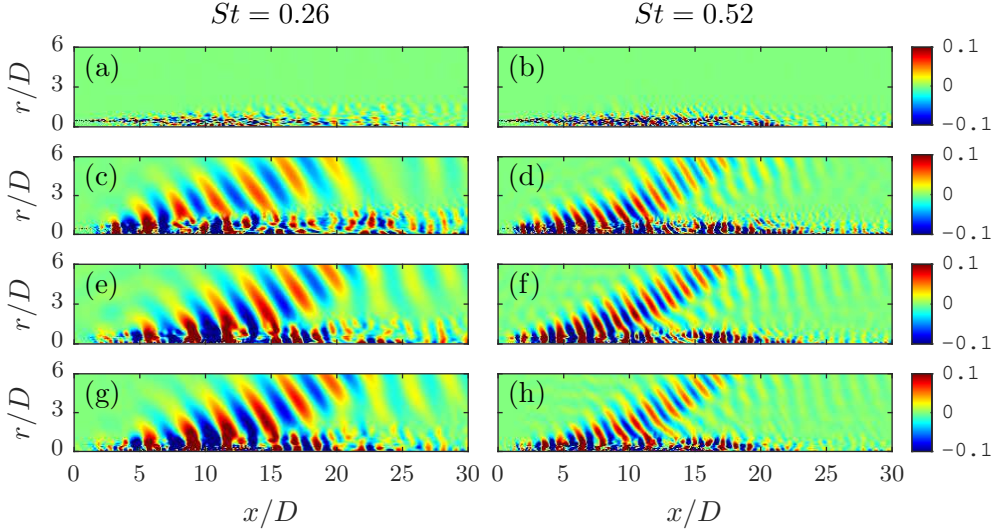}
\caption{Realizations of the Mach 1.5 jet at $St=0.26$ and $m=0$, including an LES derived forcing and pressure response realization (top two rows respectively) and LNS and OWNS computed realizations (bottom two rows respectively) forced with the above LES forcing realization. Contours for all plots are from $\pm 5 \times 10^{-4}$. }
\label{fig:Realizations}
\end{figure}

Figure \ref{fig:Realizations} shows one realization of the forcing and response for $St = 0.26$ and 0.52.  Specifically, we show the component of the forcing that is applied to the energy equation and the response of the pressure field; other components of the forcing and response are qualitatively similar \citep{Towne:2016}.  The forcing is rather incoherent and lacks readily visible large-scale structures.  In contrast, the pressure field contains distinct wavepacket structures and an acoustic beam emanating from the near-field to the far-field.  Despite the lack of structure within the forcing, its introduction as a volumetric forcing term to the LNS and OWNS-P operators produces a response that is quite similar to the LES pressure field (differences can be attributed to factors described earlier).  

Of greater relevance to the current study, the OWNS-P response closely matches the LNS response.  For both frequencies, the OWNS-P response accurately captures the relevant dynamics in the jet, including near-field structures such as Kelvin-Helmholtz wavepackets (similar to those observed in Figure 3) and Orr-type wavepackets (farther downstream, e.g., $15<x/D,20$), as well as the angle and intensity of acoustic radiation to the far field.  This substantial agreement between the LNS and OWNS-P responses indicates that the OWNS-P approximation of the action of the resolvent operator on this forcing vector accurately mimics that of the true resolvent operator.

\begin{figure}
\centering
\input{./Figures/Jet_PSD.tex}
\includegraphics[trim={0cm 0cm 0cm 0.5cm},clip]{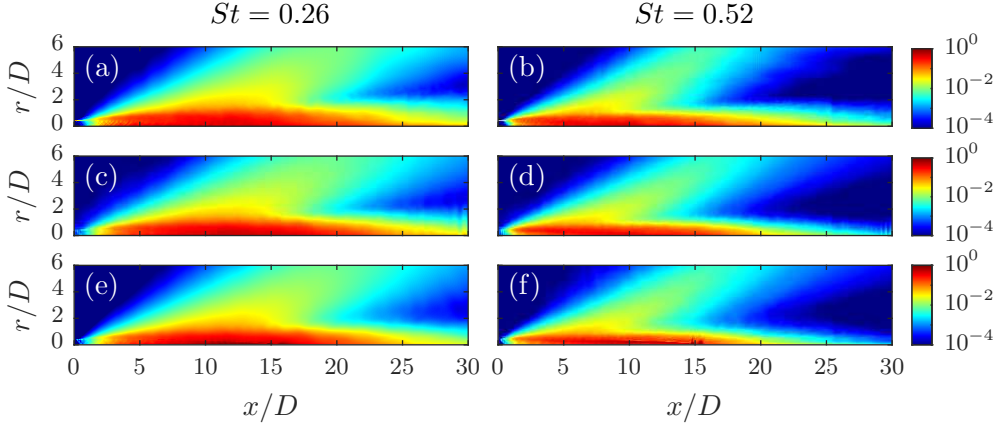}
\caption{Full-field RMS pressure results for LES (top), LNS (middle), and OWNS (bottom) at $St = 0.26$ (left) and $0.52$ (right). }
\label{fig:Jet_PSD}
\end{figure}

The ensemble of LES flow and forcing realizations can be used to compute the average error of the OWNS-P approximation over many potential forcing vectors.  Figure~\ref{fig:Jet_PSD} shows the power spectral density (PSD) of the LES, LNS, and OWNS-P pressure fields for $St = 0.26$ and 0.52, i.e., the squared-amplitude of the response averaged over the ensemble of realizations.  In each case, the PSD is scaled by the maximum value of the PSD of the LES data.  The PSD of the LNS and OWNS-P solutions accurately mimic that of the LES data.  More importantly, the PSD computed from the ensemble of OWNS-P solutions closely matches the PSD computed from the LNS solutions.  Both the envelope of the high-energy near-field region and intensity and directivity of the acoustic beam show good agreement for both frequencies.  

\begin{figure}
\centering
\input{./Figures/Jet_ErrorPSD.tex}
\includegraphics[trim={0cm 0cm 0cm 0.5cm},clip]{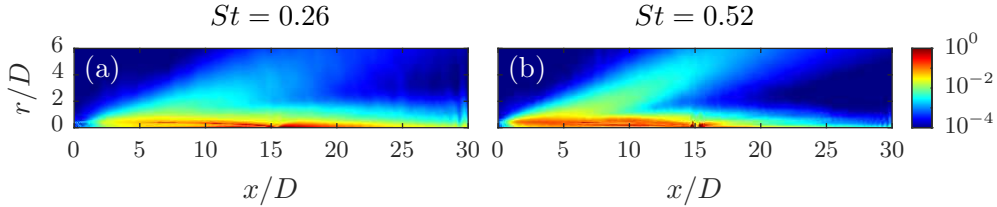}
\caption{Full-field error between LES pressure RMS and both LNS (top) and OWNS (bottom) at $St = 0.26$ (left) and $St = 0.52$ (right). }
\label{fig:Jet_errorPSD}
\end{figure}

The PSD of the error between the LNS and OWNS-P solutions, normalized by the maximum value of the PSD of the LNS solution, is shown in Figure~\ref{fig:Jet_errorPSD}.  Comparing with figure~\ref{fig:Jet_PSD}, the average error is observed to be at least an order of magnitude smaller than the average solution in both the near-field hydrodynamic region and the acoustic field.  Overall, these results indicate that the OWNS-P method provides an accurate approximation of the action of the resolvent operator for the ensemble of forcing realizations extracted from the LES data; we will see in Part 2 of the paper that it also provides an accurate approximation for the optimal forcing vectors that excite the largest response.

\section{Conclusions}
\label{Sec:Conclusions}

Computing resolvent modes for flows with multiple inhomogeneous directions remains a computationally intensive task.  In this paper, we have laid the groundwork for a new method that reduces the cost of computing resolvent modes for flows that include a slowly varying direction.  Specifically, we showed how the action of the resolvent operator on a forcing vector can be accurately approximated using a spatial marching method.  This approach constitutes an extension of the one-way Navier-Stokes methodology introduced by \cite{Towne2015oneway} to accommodate a forcing term on the linear equations, which is central to the concept of resolvent analysis.  

The method is based on an approximation of the projection operator that rigorously splits the solution vector into downstream- and upstream-traveling components.  Applying this projection operator to the linearized Navier-Stokes equations yields a well-posed equation governing the downstream evolution of the flow, which can be stably integrated in the slowly varying direction.  This projection-based method, which we call OWNS-P, uses the same recursion parameters and inherits the same convergence properties as the previous outflow-based OWNS methodology, OWNS-O, of \cite{Towne2015oneway}.  Unlike the ubiquitous parabolized stability equations, the OWNS-P method is capable of capturing the complete downstream response of the flow rather than a single instability mechanism \citep{Towne2019critical}. 

The OWNS-P method was demonstrated using an acoustic propagation problem and a turbulent jet.  For the jet, we showed that the OWNS-P solution faithfully approximates the action of the resolvent operator on both a localized forcing near the nozzle exit designed to excite the Kelvin-Helmholtz instability and a distributed volumetric forcing extracted from LES data.  In the latter case, the OWNS-P solution captured all three prominent downstream-traveling mechanisms present in the jet: Kelvin-Helmholtz, Orr, and acoustic waves.  

In part 2 of the paper, we will show how the ability to efficiently approximate the action of the resolvent operator on a forcing vector via spatial marching can be leveraged to compute global resolvent modes at substantially reduced cost.  This capability could enable computation of global resolvent modes for previously intractable three-dimensional turbulent flows that contain a slowly varying direction, including swept wings and other three-dimensional aerodynamic bodies and jets with non-circular nozzles, e.g., rectangular jets and other complex nozzle geometries typical of modern tactical aircraft.

The OWNS methodology is also applicable to other problems beyond computation of resolvent modes.  For example, OWNS could be used to find the downstream receptivity to an incident disturbance, compute kernels for closed-loop flow control, or predict flow statistics based on stochastic forcing.  Indeed, any problem to which PSE can be applied can also be addressed using OWNS, and the latter approach will provide greater accuracy for problems containing multiple relevant physical mechanisms, e.g., multiple instability modes, transient growth, or acoustics.  The proper choice between the original OWNS-O method and the OWNS-P method developed in this paper is dictated by the need, or lack-there-of, for a forcing term: only the OWNS-P method can accommodate a forcing term, but at the cost of solving a system of recursion equations that is twice as large compared to the OWNS-O approach.


\section*{Acknowledgments}
The authors thank Mr. Liam Heidt for his assistance in analyzing and measuring the FLOPS and memory requirements reported in \S~\ref{sec:OWE_projection_cost}.  AT was supported in part by a catalyst grant from the Michigan Institute for Computational Discovery and Engineering (MICDE).  GR, EP, and TC were supported by the Boeing Company through a Strategic Research and Development Relationship Agreement CT-BA-GTA-1 and by ONR grants N0014-11-1-0753, N00014-16-1-2445, and N00014-21-1-2158.


\begin{appendices}

\section{Analytical solution of the recursion equations}
\label{App:Recursions}

In this appendix, we provide an analytical solution of the recursion equations~(\ref{Eq:OWEP_recursions}), which leads to the expression for $\tc{P}_{N_{\beta}}$ given in~(\ref{Eq:OWEP_P_approx}).

We begin by writing~(\ref{Eq:OWEP_recursions_b})~-~(\ref{Eq:OWEP_recursions_d}) in terms of the expansion coefficients $\boldsymbol{\psi}^{j} = \tc{U}\boldsymbol{\hat{\phi}}_{\pm}^{j}$,
\begin{subequations}
\label{Eq:OWEP_recursionsD}
	\begin{align}
	&\left(\tc{D} - i \beta_{-}^{j} \tc{I} \right) \boldsymbol{\psi}^{-j} - \left(\tc{D} - i \beta_{+}^{j} \tc{I} \right) \boldsymbol{\psi}^{-j-1} = 0 \qquad \qquad &&j = 1,\dots,N_{\beta}-1\\
	&\left(\tc{D} - i \beta_{-}^{0} \tc{I} \right) \boldsymbol{\psi}^{0} - \left(\tc{D} - i \beta_{+}^{0} \tc{I} \right) \boldsymbol{\psi}^{-1} = \left(\tc{D} - i  \beta_{-}^{0} \tc{I} \right) \boldsymbol{\psi} \\
	&\left(\tc{D} - i \beta_{+}^{j} \tc{I} \right) \boldsymbol{\psi}^{j} - \left(\tc{D} - i \beta_{-}^{j} \tc{I} \right) \boldsymbol{\psi}^{j+1} = 0 \qquad \qquad &&j = 0,\dots,N_{\beta}-1.
	\end{align}
\end{subequations}
Since $\tc{D}$ is diagonal, each scalar component of $\boldsymbol{\psi}^{j}$ can be treated separately,
\begin{subequations}
\label{Eq:OWEP_recursionsK}
	\begin{align}
	&\left(i \alpha_{k} - i \beta_{-}^{j} \tc{I} \right) \boldsymbol{\psi}^{-j}_{k} - \left(i \alpha_{k} - i \beta_{+}^{j} \tc{I} \right) \boldsymbol{\psi}^{-j-1}_{k} = 0 \qquad \qquad &&j = 1,\dots,N_{\beta}-1\\
	&\left(i \alpha_{k} - i \beta_{-}^{0} \tc{I} \right) \boldsymbol{\psi}^{0}_{k} - \left(i \alpha_{k} - i \beta_{+}^{0} \tc{I} \right) \boldsymbol{\psi}^{-1}_{k} = \left(i \alpha_{k} - i  \beta_{-}^{0} \tc{I} \right) \boldsymbol{\psi}_{k} \\
	&\left(i \alpha_{k} - i \beta_{+}^{j} \tc{I} \right) \boldsymbol{\psi}^{j}_{k} - \left(i \alpha_{k} - i \beta_{-}^{j} \tc{I} \right) \boldsymbol{\psi}^{j+1}_{k} = 0 \qquad \qquad &&j = 0,\dots,N_{\beta}-1
	\end{align}
\end{subequations}
for $k = 1,\dots,N$. The intermediate auxiliary variables ($j = -N_{\beta}+1,\dots,N_{\beta}-1$) can then be easily eliminated, leaving
\begin{subequations}
\label{Eq:OWEP_RecSolved_f}
	\begin{align}
	&\boldsymbol{\psi}_{k}^{-N_{\beta}} = \mathcal{F}^{-1}(\alpha_{k}) \left( \boldsymbol{\psi}_{k}^{0} - \boldsymbol{\psi}_{k}\right), \\
	&\boldsymbol{\psi}_{k}^{N_{\beta}} = \mathcal{F}(\alpha_{k}) \boldsymbol{\psi}_{k}^{0}.
	\end{align}
\end{subequations}
The function $\mathcal{F}\left(\alpha \right)$ is the same as in~(\ref{Eq:ApproxParDAErecursionR}).  Equation~(\ref{Eq:OWEP_RecSolved_F}) can be written for all $k$ as
\begin{subequations}
\label{Eq:OWEP_RecSolved_F}
\begin{align}
&\left\{\begin{array}{c} \boldsymbol{\psi}_{+}^{-N_{\beta}} \\ \boldsymbol{\psi}_{-}^{-N_{\beta}} \end{array}\right\} = \left[\begin{array}{cc} \tc{F}_{++}^{-1} & \tc{0} \\ \tc{0} & \tc{F}_{--}^{-1} \end{array}\right]\left\{\begin{array}{c} \boldsymbol{\psi}_{+}^{0} - \boldsymbol{\psi}_{+} \\ \boldsymbol{\psi}_{-}^{0}-\boldsymbol{\psi}_{-} \end{array}\right\}, \\
&\left\{\begin{array}{c} \boldsymbol{\psi}_{+}^{N_{\beta}} \\ \boldsymbol{\psi}_{-}^{N_{\beta}} \end{array}\right\} = \left[\begin{array}{cc} \tc{F}_{++} & \tc{0} \\ \tc{0} & \tc{F}_{--} \end{array}\right]\left\{\begin{array}{c} \boldsymbol{\psi}_{+}^{0} \\ \boldsymbol{\psi}_{-}^{0} \end{array}\right\},
\end{align}
\end{subequations}
where, $\tc{F}_{++}$ and $\tc{F}_{--}$ are diagonal matrices whose entries are the values of $\mathcal{F}(\alpha)$ associated with each downstream- and upstream-traveling eigenvalue, respectively.

To apply the termination conditions given by~(\ref{Eq:OWEP_recursions_a})~and~(\ref{Eq:OWEP_recursions_e}), it is necessary to write the left-hand-sides of~(\ref{Eq:OWEP_RecSolved_F}) in terms of $\boldsymbol{\hat{\phi}}_{\pm}^{N_{\beta}}$ and $\boldsymbol{\hat{\phi}}_{\pm}^{-N_{\beta}}$.  This requires a further parting further partitioning of the left eigenvectors of $\tc{M}$,
\begin{equation}
\label{Eq:eigenvector_compare}
 \tc{U}_{+} = \left[ \begin{array}{cc} \tc{U}_{++} & \tc{U}_{+-} \end{array} \right] \quad \tc{U}_{-} = \left[ \begin{array}{cc} \tc{U}_{-+} & \tc{U}_{--} \end{array} \right],
\end{equation}
where $\tc{U}_{+-} \in \mathbb{C}^{N_{+} \times N_{-}}$ and so on.
Then, (\ref{Eq:OWEP_RecSolved_F}) can be written as
\begin{subequations}
\label{Eq:OWEP_ApproxParDAErecusrionSolvedR2}
\begin{align} 
&\left[\begin{array}{cc} \tc{U}_{++} & \tc{U}_{+-} \\ \tc{U}_{-+} & \tc{U}_{--} \end{array}\right]\left\{\begin{array}{c} \boldsymbol{\hat{\phi}}_{+}^{-N_{\beta}} \\ \boldsymbol{\hat{\phi}}_{-}^{-N_{\beta}} \end{array}\right\} = \left[\begin{array}{cc} \tc{F}_{++}^{-1} & \tc{0} \\ \tc{0} & \tc{F}_{--}^{-1} \end{array}\right]\left\{\begin{array}{c} \boldsymbol{\psi}_{+}^{0} - \boldsymbol{\psi}_{+} \\ \boldsymbol{\psi}_{-}^{0}-\boldsymbol{\psi}_{-} \end{array}\right\}, \\
&\left[\begin{array}{cc} \tc{U}_{++} & \tc{U}_{+-} \\ \tc{U}_{-+} & \tc{U}_{--} \end{array}\right]\left\{\begin{array}{c} \boldsymbol{\hat{\phi}}_{+}^{N_{\beta}} \\ \boldsymbol{\hat{\phi}}_{-}^{N_{\beta}} \end{array}\right\} = \left[\begin{array}{cc} \tc{F}_{++} & \tc{0} \\ \tc{0} & \tc{F}_{--} \end{array}\right]\left\{\begin{array}{c} \boldsymbol{\psi}_{+}^{0} \\ \boldsymbol{\psi}_{-}^{0} \end{array}\right\}.
\end{align}
\end{subequations}
Applying equations~(\ref{Eq:OWEP_recursions_a})~and~(\ref{Eq:OWEP_recursions_e}) and eliminating $\boldsymbol{\hat{\phi}}_{-}^{-N_{\beta}}$ and $\boldsymbol{\hat{\phi}}_{+}^{N_{\beta}}$ leaves
\begin{equation}
\label{Eq:OWEP_psiProj}
\left[\begin{array}{cc} \tc{I} & -\tc{R}_{+-} \\ -\tc{R}_{-+} & \tc{I} \end{array}\right]\left\{\begin{array}{c} \boldsymbol{\psi}_{+}^{0} \\ \boldsymbol{\psi}_{-}^{0} \end{array}\right\} =\left[\begin{array}{cc} \,\,\,\tc{I}\,\,\, & \tc{0} \\ \tc{0} & \,\,\,\tc{0}\,\,\, \end{array}\right] \left[\begin{array}{cc} \tc{I} & -\tc{R}_{+-} \\ -\tc{R}_{-+} & \tc{I} \end{array}\right]\left\{\begin{array}{c} \boldsymbol{\psi}_{+} \\ \boldsymbol{\psi}_{-} \end{array}\right\}
\end{equation}
with
\begin{subequations}
\label{Eq:OWEP_R}
\begin{align}
&\tc{R}_{+-} = \tc{F}_{++} \tc{W_{+-}} \tc{F}_{--}^{-1}, \\
&\tc{R}_{-+} = \tc{F}_{--}^{-1} \tc{W_{-+}} \tc{F}_{++},
\end{align}
\end{subequations}
and 
\begin{subequations}
\label{Eq:OWEP_W}
\begin{align}
&\tc{W}_{+-} = \tc{U}_{+-} \tc{U}_{--}^{-1}, \\
&\tc{W}_{-+} = \tc{U}_{-+} \tc{U}_{++}^{-1}.
\end{align}
\end{subequations}
Solving equation~(\ref{Eq:OWEP_psiProj}) for $\boldsymbol{\psi}^{0}$ and reverting to $\boldsymbol{\hat{\phi}}_{\pm}$ variables gives
\begin{equation}
\label{OWEP_phiprime_approx}
\boldsymbol{\hat{\phi}}_{\pm}^{0} = \tc{P}_{N_{\beta}} \boldsymbol{\hat{\phi}}_{\pm},
\end{equation}
where $\tc{P}_{N_{\beta}}$ is given by~(\ref{Eq:OWEP_P_approx}).

\section{Well-posedness of the approximate one-way equation}
\label{App:WellPosed}

To show that~(\ref{Eq:OWE-P1})~and~(\ref{Eq:OWE-P2}) remain well-posed as one-way equations when $\tc{P}_{N_{\beta}}$ is used in place of $\tc{P}$, it suffices to show that $\tc{P}_{N_{\beta}} \to \tc{P}$ as $\eta \to \infty$, since we know that the eigenvalues of $\tc{P}\tc{M}$ and $\left(\tc{I}-\tc{P}\right)\tc{M}$ exhibit the correct behavior for well-posedness, as defined by~(\ref{Eq:rightgoing_X})~and~(\ref{Eq:leftgoing_X}), in this limit.  Therefore, the task at hand is to show that $\tc{R}_{+-}$ and $\tc{R}_{-+}$ go to zero as $\eta \to \infty$.  In this limit, $\tc{M}$ tends to the diagonal matrix $-\eta\tilde{\tc{A}}^{-1}$.  Since it is diagonal, its eigenvector matrices are also diagonal and unitary; $\tc{U}_{++}$ and $\tc{U}_{--}$ are appropriate sized identity matrices and $\tc{U}_{+-}$ and $\tc{U}_{-+}$ are zero.  Also, the eigenvalues of the asymptotic form of $\tc{M}$ approach infinity, which causes $\mathcal{F}\left( \alpha_{k} \right) \to 1$ for every $k$ since the recursion parameters are bounded as $\eta \to \infty$ \citep{Towne2015oneway}.  Consequently, $\boldsymbol{F}_{++}$ and $\boldsymbol{F}_{--}$ become identity matrices.  Putting this all together, we conclude from~(\ref{Eq:OWEP_R}) that $\tc{R}_{+-}$ and $\tc{R}_{-+}$ go to zero as $\eta \to \infty$.  Therefore, (\ref{Eq:OWE-P1})~and~(\ref{Eq:OWE-P2}) are well-posed when $\tc{P}_{N_{\beta}}$ is used in place of $\tc{P}$.

\section{Approximate projection in matrix form}
\label{App:OWNS_mats}



The recursions~(\ref{Eq:OWEP_recursions}) and~(\ref{Eq:ApproxPhiPrime}) define a system of equations whose solution provides the action of the approximate projection operator on a vector.  These equations are written in terms of $\tc{M}$.  While this aided the theoretical development of the method, in practice it is preferable to work in terms of $\tc{L}$ and $\tilde{\tc{A}}$.  This can be achieved by introducing zero characteristic components to the auxiliary variables satisfying
\begin{equation}
\label{Eq:aux_0}
\boldsymbol{\hat{\phi}}_{0}^{j} = - \tc{L}_{00}^{-1} \tc{L}_{0\pm} \boldsymbol{\hat{\phi}}_{\pm}^{j}.
\end{equation}
Then, (\ref{Eq:OWEP_recursions}), (\ref{Eq:ApproxPhiPrime}), and~(\ref{Eq:aux_0}) can be written together in the form shown in~(\ref{Eq:OWEP_recursions_system}).  Here, we exhibit the structure of this system using the example $N_{\beta} = 2$:

\small
\begin{equation}
\boldsymbol{\hat{\phi}}_{\pm}' \approx \boldsymbol{\hat{\phi}}_{\pm}^{0} = 
\underbrace{\begin{bmatrix} 
\left[\tc{0}_{\pm-} \,\, \tc{0}_{\pm0} \right] & \tc{0} & \left[\tc{I}_{\pm\pm} \,\, \tc{0}_{\pm0} \right] & \tc{0} & \left[\tc{0}_{\pm+} \,\, \tc{0}_{\pm0} \right]
\end{bmatrix} }_{\tc{P}_{3}}
\renewcommand\arraystretch{1.5}
\underbrace{\begin{bmatrix}
\begin{bmatrix}\boldsymbol{\hat{\phi}}_{-}^{-2} \\ \boldsymbol{\hat{\phi}}_{0}^{-2} \end{bmatrix}	\\ 
\boldsymbol{\hat{\phi}}^{-1} 		\\ 
\boldsymbol{\hat{\phi}}^0  		    \\ 
\boldsymbol{\hat{\phi}}^{1} 	    \\ 
\begin{bmatrix}\boldsymbol{\hat{\phi}}_{+}^{2} \\ \boldsymbol{\hat{\phi}}_{0}^{2} \end{bmatrix}
\end{bmatrix}}_{\boldsymbol{\hat{\phi}}^{\mathrm{aux}}},
\end{equation}

\begin{align}
\label{Eq:OWNS_block_matrix}
\renewcommand\arraystretch{1.75}
\underbrace{\left[
\begin{array}{c|c|c|c|c}
-\begin{bmatrix}\tc{Q}_{-}^{(+1)} \,\, \tc{Q}_{0}^{(+1)} \end{bmatrix} & \tc{M}^{(-1)} &  &  &  \\ \hdashline 
		& -\tc{Q}^{(+0)} 	& \tc{Q}^{(-0)} & \ \\ \hdashline 
		& 	& \begin{bmatrix}\tc{L}_{0\pm} \,\, \tc{L}_{00}  \end{bmatrix} & \ \\ \hdashline 
 		& 			& -\tc{Q}^{(+0)}	& \tc{Q}^{(-0)} \\ \hdashline 
	&  			&  			& -\tc{Q}^{(+1)} 	& 	\begin{bmatrix}\tc{Q}_{+}^{(-1)} \,\, \tc{Q}_{0}^{(-1)}  \end{bmatrix}
\end{array}
\right]}_{\tc{P}_{2}}
\renewcommand\arraystretch{1.5}
\underbrace{\begin{bmatrix}
\begin{bmatrix}\boldsymbol{\hat{\phi}}_{-}^{-2} \\ \boldsymbol{\hat{\phi}}_{0}^{-2} \end{bmatrix}	\\ 
\boldsymbol{\hat{\phi}}^{-1} 		\\ 
\boldsymbol{\hat{\phi}}^0  		    \\ 
\boldsymbol{\hat{\phi}}^{1} 	    \\ 
\begin{bmatrix}\boldsymbol{\hat{\phi}}_{+}^{2} \\ \boldsymbol{\hat{\phi}}_{0}^{2} \end{bmatrix}
\end{bmatrix}}_{\boldsymbol{\hat{\phi}}^{\mathrm{aux}}} \nonumber \\
= 
\renewcommand\arraystretch{1.75}
\underbrace{\begin{bmatrix}
\tc{0} \\  \tc{Q}^{(-0)} \\ \tc{0} \\ \tc{0} \\ \tc{0}
\end{bmatrix} \begin{bmatrix} \tc{I} \\ -\tc{L}_{00}^{-1} \tc{L}_{0\pm} \ \end{bmatrix}}_{\tc{P}_{1}} \boldsymbol{\hat{\phi}}_{\pm},
%
\end{align}
\normalsize
where 
\begin{subequations}
\label{Eq:Q_def}
\begin{align}
\tc{Q}^{(+j)} &= \tc{L} - i \beta_{+}^{j} \tilde{\tc{A}}, \\
\tc{Q}^{(-j)} &= \tc{L} - i \beta_{-}^{j} \tilde{\tc{A}}.
\end{align}
\end{subequations}
The superscripts in~(\ref{Eq:Q_def}) indicates which recursion parameter is subtracted, not a power of the matrix.  The single plus, minus, and zero matrix subscripts indicate certain columns of a matrix, as in~(\ref{Eq:M_eig_decomp}).  For example, $\tc{Q}^{(-1)}_{+}$ contains the first $N_{+}$ columns of $\tc{Q}^{(-1)}$.  These truncated blocks appear in~(\ref{Eq:OWNS_block_matrix}) due to the terminal conditions~(\ref{Eq:OWEP_recursions_a}) and~(\ref{Eq:OWEP_recursions_e}) within the recursion equations, which are enforced by removing rows and columns corresponding to $\boldsymbol{\hat{\phi}}_{-}^{N_{\beta}}$ and $\boldsymbol{\hat{\phi}}_{+}^{-N_{\beta}}$.  Indeed, it is these terminal conditions, and their impact on $\tc{P}_{2}$, that force the recursions to be solved all at once as a coupled set.

\section{Linearized Navier-Stokes operators}
\label{App:NS_ops}

Starting from~(\ref{Eq:NS_eqs}), the linearized Navier-Stokes equations in two-dimensional Cartesian coordinates can be written in the form of~(\ref{Eq:LinearPDE_f}) with 
\begin{equation}
\label{Eq:NS_2D_B_def}
\renewcommand\arraystretch{1.2}
\mathcal{B} = \mathcal{B}_{y} \frac{\partial}{\partial y} + \mathcal{B}_{yy} \frac{\partial}{\partial y^{2}} + \mathcal{B}_{0},
\end{equation}

\begin{equation*}
\renewcommand\arraystretch{1.2}
\mathcal{A} =\left[\begin{array}{cccc} \bar{u}_{x} & -\bar{\varv}&0 & 0 \\ 0 &  \bar{u}_{x} & 0  & \bar{\varv} \\ 0 & 0 & \bar{u}_{x} & 0 \\ 0 & \gamma\bar{p} & 0 & \bar{u}_{x} \end{array}\right], \qquad \mathcal{B}_{0} = \left[\begin{array}{cccc} - \nabla \cdot \bar{u} & \partderiv{\bar{\varv}}{x} & \partderiv{\bar{\varv}}{y} & 0 \\ \partderiv{\bar p}{x} - \frac{1}{Re}\nabla^{2}\bar{u}_{x} & \partderiv{\bar u_x}{x} & \partderiv{\bar u_x}{y} & 0 \\\partderiv{\bar p}{y} - \frac{1}{Re}\nabla^{2}\bar{u}_{y} & \partderiv{\bar u_{y}}{x} & \partderiv{\bar u_{y}}{y} & 0 \\ - \frac{\gamma}{Re Pr}\nabla^{2}\bar{p} & \partderiv{\bar p}{x} & \partderiv{\bar p}{y}  & \gamma \nabla \cdot \bar{u} - \frac{\gamma}{Re Pr}\nabla^{2}\bar{\varv} \end{array}\right], 
\end{equation*}

\begin{equation*}
\renewcommand\arraystretch{1.2}
\mathcal{B}_{y} =\left[\begin{array}{cccc} \bar{u}_{y} & 0 & -\bar{\varv} & 0 \\ 0 &  \bar{u}_{y} & 0 & 0 \\ 0 & 0 & \bar{u}_{y} & \bar{\varv} \\ 0 & 0 & \gamma\bar{p} & \bar{u}_{y}  \end{array}\right], \qquad \mathcal{B}_{yy} = -\frac{1}{Re} \left[\begin{array}{cccc} 0 & 0 & 0 & 0 \\ 0 &  \bar{\varv} & 0 & 0  \\ 0 & 0 & \bar{\varv} & 0 \\ \gamma\bar{p}/Pr & 0 & 0 & \gamma\bar{\varv}/Pr \end{array}\right],
\end{equation*}
and $q = \left\{ \varv, u_{x}, u_{y}, p \right\}^T$.  Second $x$-derivatives have been neglected as discussed in \S~\ref{Sec:RightVsLeft}.  Similar expressions can be obtained for other coordinate systems, e.g., three-dimensional Cartesian or cylindrical coordinates, but are omitted for brevity.  

\end{appendices}

\bibliographystyle{jfm}
\bibliography{bib_OWNSP}

\end{document}